\documentclass[preprint,sort&compress,12pt]{elsarticle}

\usepackage{graphicx}
\usepackage{amsmath}

\usepackage{amssymb}

\usepackage{amsthm}
\usepackage{url}

\usepackage{subfig}
\usepackage{enumerate}
\usepackage{fullpage}
\usepackage{algorithm}
\usepackage{algpseudocode}
\usepackage{xcolor}
\usepackage{enumitem}
\usepackage{mathtools}

\newcommand{\bs}[1]{\boldsymbol{#1}}

\usepackage{soul}
\usepackage{array}
\newcolumntype{M}[1]{>{\centering\arraybackslash}m{#1}}

\journal{XXX}
\begin{document}

\begin{frontmatter}


  \title{Quantification of Uncertainties in Turbulence Modeling: A Comparison of Physics-Based and Random
    Matrix Theoretic Approaches}




\author{Jian-Xun Wang}
\author{Rui Sun}
\author{Heng Xiao\corref{corxh}}
\cortext[corxh]{Corresponding author. Tel: +1 540 231 0926}
\ead{hengxiao@vt.edu}

\address{Department of Aerospace and Ocean Engineering, Virginia Tech, Blacksburg, VA 24060, United States}

\begin{abstract}
  Numerical models based on Reynolds-Averaged Navier-Stokes (RANS) equations are widely used in
  engineering turbulence modeling. However, the RANS predictions have large model-form uncertainties
  for many complex flows, e.g., those with non-parallel shear layer or strong mean flow
  curvature. Quantification of these large uncertainties originating from the modeled Reynolds
  stresses has attracted attention in turbulence modeling community.  Recently, a physics-based
  Bayesian framework for quantifying model-form uncertainties has been proposed with successful
  applications to several flows.  Nonetheless, how to specify proper priors without introducing
  unwarranted, artificial information remains challenging to the current form of the physics-based
  approach. Another recently proposed method based on random matrix theory provides the prior
  distributions with the maximum entropy, which is an alternative for model-form uncertainty
  quantification in RANS simulations.  This method has better mathematical rigorousness and provides
  the most non-committal prior distributions without introducing artificial constraints. On the
  other hand, the physics-based approach has the advantages of being more flexible to incorporate
  available physical insights.  In this work, we utilize the random matrix theoretic approach to
  assess and possibly improve the specification of priors used in the physics-based approach.  A
  comparison of the two approaches is then conducted through a test case using a canonical flow, the
  flow past periodic hills. The numerical results show that, to achieve maximum entropy in the prior
  of Reynolds stresses, the perturbations of shape parameters in Barycentric coordinates are
  normally distributed. Moreover, the perturbations of the turbulence kinetic energy should conform
  to log-normal distributions. Finally, it sheds light on how large the variance of each physical
  variable should be compared with each other to achieve the approximate maximum entropy prior.  The
  conclusion can be used as a guidance for specifying proper priors in the physics-based, Bayesian
  uncertainty quantification framework.

\end{abstract}

\end{frontmatter}

\section*{Highlights}
\begin{enumerate}
\item Compared physics-based and random matrix methods to quantify RANS model uncertainty
\item Demonstrated applications of both methods in channel flow over periodic hills
\item Examined the amount of information introduced in the physics-based approach
\item Discussed implications to modeling turbulence in both near-wall and separated regions
\end{enumerate}


\section*{Notations}
We summarize the convention of notations below because of the large number of symbols used in this
paper. The general conventions are as follows:
\begin{enumerate} 
\item Upper case letters with brackets (e.g., $[R]$) indicate matrices or tensors; lower case
  letters with arrows (e.g., $\vec{v}$) indicate vectors; undecorated letters in either upper or
  lower cases indicate scalars.  An exception is the spatial coordinate, which is denoted as $x$ for
  simplicity but is in fact a $3 \times 1$ vector.  Tensors (matrices) and vectors are also
  indicated with index notations, e.g., $R_{ij}$ and $v_i$ with $i, j = 1, 2, 3$. 
 
\item Bold letters (e.g., $[\mathbf{R}]$) indicate random variables (including scalars, vectors, and
  matrices), the non-bold letters (e.g., $[R]$) indicate the corresponding realizations, and
  underlined letters (e.g., $[\underline{R}]$) indicate the mean.

\item Symbols ${M}_d^{+}$, and $\mathbb{M}_d^{+0}$ indicate the sets of
  symmetric positive definite and symmetric positive semi-definite matrices,
  respectively, of dimension $d \times d$ with the following relation: $\mathbb{M}_d^{s} \subset
  {M}_d^{+} \subset \mathbb{M}_d^{+0}$.
\end{enumerate}
This work deals with Reynolds stresses, which are rank two tensors. Therefore, it is implied
throughout the paper that all random or deterministic matrices have sizes $3 \times 3$ with real
entries unless noted otherwise.  Finally, a list of nomenclature is presented in~\ref{app:notation}.

\section{Introduction}
Despite the increasing availability of computational resources in the past decades, high-fidelity
simulations (e.g., large eddy simulation, direct numerical simulation) are still not affordable for
most practical problems.  Numerical models based on Reynolds-Averaged Navier--Stokes (RANS)
equations are still the dominant tools for the prediction of turbulent flows in industrial and
natural processes.  However, for many practical flows, e.g., those with strong adverse pressure
gradient, non-parallel shear layer, or strong mean flow curvature, the predictions of RANS models
have large uncertainties. The uncertainties are mostly attributed to the phenomenological closure
models for the Reynolds stresses~\cite{pope2000turbulent,oliver2009uncertainty}. Previous efforts in
quantifying and reducing model-form uncertainties in RANS simulations have mostly followed
parametric approaches, e.g., by perturbing, tuning, or inferring the parameters of the closure
models of the Reynolds stress~\cite{margheri2014epistemic, edeling2014bayesian, edeling2014predictive}.

Recently, the turbulence modeling community has recognized the limitations of the parametric
approaches and started investigating non-parametric approaches where uncertainties are directly
injected into the Reynolds
stresses~\cite{oliver2009uncertainty,emory2011modeling,dow2011quantification,emory2013modeling,gorle2013framework,xiao-mfu}.
In their pioneering work, Iaccarino et
al.~\cite{emory2011modeling,emory2013modeling,gorle2013framework} proposed a physics-based approach,
where the Reynolds stress is projected onto six physically meaningful dimensions (its
shape, magnitude, and orientation).  They further perturbed the Reynolds stresses towards the
limiting states in the physically realizable range, based on which the RANS prediction uncertainties
are estimated.  Building on the work of Iaccarino et
al.~\cite{emory2011modeling,emory2013modeling,gorle2013framework}, Xiao et al.~\cite{xiao-mfu}
modeled the Reynolds stress discrepancy as a \emph{zero-mean random field} and
used a physical-based parameterization to systematically explore the uncertainty space.  
They further used Bayesian inferences to incorporate observation data to reduce the
model-form uncertainty in RANS simulation. While the physics-based method has achieved
significant successes, the method in its current form has two major limitations. First, 
uncertainties are only injected to the shape and magnitude of the Reynolds stresses but not to the
orientations, and thus they do not fully explore the uncertainty space.  Second, it is challenging
to specify prior distributions over these physical variables without introducing artificial constraints.  
The priors are critical for uncertainty propagation and Bayesian
inference, particularly when the amount of data is limited~\cite{wang2015incorporating}. Xiao et
al.~\cite{xiao-mfu} specified Gaussian distribution for the perturbations of shape parameters in
natural coordinates and log-normal distribution for the turbulence kinetic energy
discrepancy. The perturbations in all physical parameters share the same variance field.
However, it is not clear if or how much artificial constraints are introduced into the prior with
this choice.  Moreover, without sufficient physical insight, it is not clear how large the
variance of perturbation for each physical variable should be relative to each other.

In information theory, Shannon entropy is an important measure of the information contained in each
probability distribution. The distribution best representing the current state is the one with the
largest information entropy, which is known as principle of maximum
entropy~\cite{guiasu1985principle}.  This principle has been used as a guideline to specify prior
distributions in Bayesian framework~\cite{jaynes1957information}. Although this theory has been
extensively used in information processing problems such as communications and image processing, the
application in conjunction with random matrix theory applied to physical systems is only a recent
development, which was first proposed and developed by Soize et al.~\cite{soize2000nonparametric,
  das2009bounded}.  Built on the theories developed by Soize et al., Xiao et al.~\cite{xiao-mfu4}
proposed a random matrix theoretic (RMT) approach with maximum entropy principle to quantify
model-form uncertainties in RANS simulations.  The RMT approach is an alternative to the physics-based
approach in quantifying model-form uncertainties in RANS simulations. It can provide objective priors for
Bayesian inferences that satisfies the given constraints without introducing artificial information.

While the RMT approach has better mathematical rigorousness and provides a
proper prior of the Reynolds stress tensors with maximum entropy, it has its own limitations. In
particular, since the perturbations are directly introduced to the Reynolds stress itself, it is not
straightforward to incorporate physical insights that are available for specific flows into the RMT
approach.  For example, for the flow in a channel with square cross section, the discrepancies of
RANS-predicted Reynolds stress mainly come from the shape of the Reynolds stress tensor, while the
predicted turbulence kinetic energy is rather accurate~\cite{wu2015bayesian}.  In this case, the perturbation
variances of shape parameters should be specified much larger than that of the turbulence kinetic
energy. Nonetheless, this piece of information is difficult to incorporate into the RMT approach. 
In comparison, the physics-based approach is more flexible and thus may be preferred
in engineering applications for both uncertainty quantification and Bayesian inferences. Therefore, the
objective of this work is to utilize the RMT approach to assess and improve the specification of
priors used in the physics-based approach.  To this end, the Reynolds stress samples with maximum
entropy distribution obtained in the RMT approach are first projected onto the physically meaningful
dimensions. Then, the distributions in the six physical dimensions are used to gauge the priors
specified in the physics-based approach. The comparison between the two approaches allows for
quantification of the amount of information introduced in the physics-based priors. It also sheds light
on the specification of appropriate prior for each physical variable when no further physical
knowledge is available.

The rest of the paper is organized as follows. Section~\ref{sec:Meth} introduces the physics-based
and RMT approaches for RANS model-form uncertainty quantification.
Section~\ref{sec:res} uses the flow over periodic hills as an example to perform the comparison
between the two approaches. The results are then presented and discussed.
Finally, Section~\ref{sec:con} concludes the paper.

\section{Comparison of Physics Based Approach and RMT Approach}

This work examines and compares two approaches, the physics-based approach and the 
random matrix theoretic approach, for quantifying RANS model-form uncertainties. We first briefly introduce
the general background of RANS-based turbulence modeling and the common assumptions of the two
approaches before presenting the technical details of the two approaches.

Reynolds averaged Navier--Stokes equations describe the mean quantities (e.g., velocity and
pressure) of the turbulent flows. They are obtained by performing time- or ensemble-averaging on the
Navier--Stokes equations, which describes the instantaneous flow quantities.  The averaging process
leads to a covariance term of the instantaneous velocities, which is referred to as Reynolds
stresses and needs model closure in RANS simulations. It is the consensus of the turbulence modeling
community that the modeling of the Reynolds stresses accounts for majority of the model-form
uncertainty in RANS simulations~\cite{pope2000turbulent}. In the physics based approach proposed by
Iaccarino et al.\cite{emory2011modeling,emory2013modeling} and further extended by Xiao et
al.~\cite{xiao-mfu,wang2015incorporating}, perturbations are directly injected to the RANS-predicted
Reynolds stresses.  Specifically, the physically meaningful projections of the Reynolds stress,
i.e., its magnitude, shape, and orientation are jointly perturbed around their respective mean
values obtained in the RANS simulation, and the perturbations are then propagated through RANS
solvers to the Quantities of Interests (QoI, e.g., velocities). The obtained ensemble is then used
to assess the uncertainties in the RANS predictions. The specific form of perturbation for each
variable is a modeling choice made by the user. The scheme ensures realizability (positive
semi-definiteness) of the Reynolds stresses.  Similar to the physics-based approach, zero-mean
perturbations are also injected to the RANS-predicted Reynolds stresses in the RMT approach with
realizability guaranteed.  In contrast to the physics-based approach, however, in the RMT approach
the true Reynolds stress is modeled as random matrices, for which a maximum entropy probabilistic
distribution is constructed under the constraint that the mean is the RANS-predicted value. The
maximum entropy distribution is then sampled to be obtain the perturbed Reynolds stress fields.

In summary, both the physics-based and the RMT approaches introduce zero-mean perturbations to the RANS
predicted Reynolds stresses, which are then propagated to the QoIs to assess
RANS prediction uncertainties. They differ in how the perturbations are introduced. The RMT
approach introduces perturbations directly in the Reynolds stress tensor and the perturbations
conform to a maximum entropy distribution, while the physics-based approach introduces perturbations
to the physically meaningful projections of the Reynolds stresses. The details of the two schemes
are presented below.

\label{sec:Meth}
\subsection{Physics-Based Approach}
\label{sec:Phy}
\begin{figure}[!htbp]
  \centering
   \includegraphics[width=0.9\textwidth]{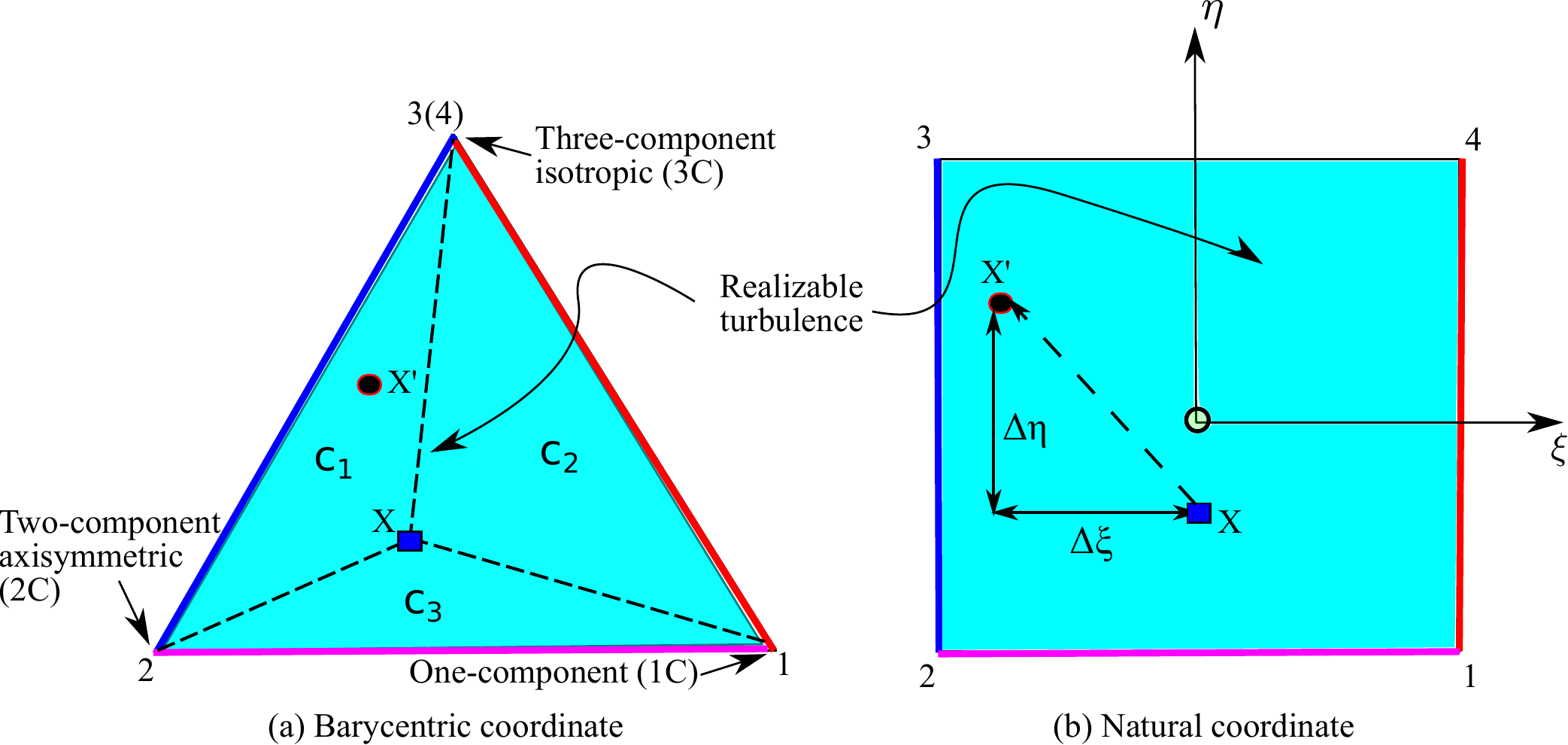}
   \caption{ Mapping between the Barycentric coordinate to the natural coordinate, transforming the
     Barycentric triangle enclosing all physically realizable
     states~\cite{banerjee2007presentation, emory2013modeling} to a square through standard finite
     element shape functions. Details of the mapping can be found in the appendix of
     ref.~\cite{xiao-mfu}. Corresponding edges in the two coordinates are indicated with matching
     colors.}
  \label{fig:bary}
\end{figure}

Here, we briefly summarize the physics-based model-form uncertainty quantification framework proposed 
by Xiao et al.~\cite{xiao-mfu} and its extension to account for uncertainties in tensor orientation.  
In the framework, the true Reynolds 
stress $[\mathbf{R}(x)]$ is modeled as a random tensorial field with the RANS-predicted 
Reynolds stress $[R(x)]^{rans}$ as prior mean, in which $x$ denotes the spatial coordinate. 
To inject uncertainties into the physically meaningful projections of Reynolds stress tensor, the 
following eigen-decomposition is performed for its each realization at any given location $x$:
\begin{equation}
  \label{eq:tau-decomp}
  [R] = 2 k \left( \frac{1}{3} [I] +  [A] \right)
  = 2 k \left( \frac{1}{3} [I] + [E] [\Lambda] [E]^T \right)
\end{equation}
where $k$ is the turbulent kinetic energy indicating the magnitude of $[R]$; $[I]$ is the
second order identity tensor; $[A]$ is the anisotropy tensor; $[E] = [\vec{e}_1, \vec{e}_2,
\vec{e}_3]$ and $[\Lambda] = \textrm{diag}[\tilde{\lambda}_1, \tilde{\lambda}_2, \tilde{\lambda}_3]$
,where $\tilde{\lambda}_1+ \tilde{\lambda}_2 + \tilde{\lambda}_3=0$, are the orthonormal eigenvectors
and the corresponding eigenvalues of $[A]$, respectively, indicating the orientation and shape of
$[R]$. In order to physically interpret the shape of the Reynolds stress and easily impose 
the realizability constraint,  the eigenvalues $\tilde{\lambda}_1$, $\tilde{\lambda}_2$, 
and $\tilde{\lambda}_3$ are mapped to the Barycentric coordinates $(C_1,C_2, C_3)$ 
with $C_1 + C_2 + C_3 = 1$. The Barycentric coordinates are defined as, 
\begin{subequations}
  \label{eq:lambda2c}
\begin{align}
  C_1 & = \tilde{\lambda}_1 - \tilde{\lambda}_2 \\
  C_2 & = 2(\tilde{\lambda}_2 - \tilde{\lambda}_3) \\
  C_3 & = 3 \tilde{\lambda}_3 + 1 \ .
\end{align}  
\end{subequations}
As shown in Fig.~\ref{fig:bary}a, the Barycentric coordinates ($C_1$, $C_2$, $C_3$) of a point
indicate the portion of areas of three sub-triangles formed by the point and with edge labeled as
$C_1$, $C_2$, and $C_3$, in the Barycentric triangle.  For example, the ratio of
the sub-triangle labeled with $C_3$ to the entire triangle is $C_3$. A point located on the top vertex
corresponds to $C_3 = 1$ while a point located on the bottom edge has $C_3$ = 0.  
The Barycentric coordinates have clear physical interpretation, i.e., the dimensionality of the turbulence.
All the edges and vertices indicate limiting states of turbulence, which are shown in Fig~\ref{fig:bary}a.
The projections of all Reynolds stresses should fall inside the Barycentric triangle to ensure the realizability. 
To facilitate parameterization, the Barycentric coordinates are further transformed to the 
natural coordinates $(\xi, \eta)$ with the triangle mapped to the square, as shown in Fig.~\ref{fig:bary}b.  
Details of the mapping can be found in ref.~\cite{xiao-mfu}. Although the tensor orientation 
($[E] = [\vec{e}_1, \vec{e}_2, \vec{e}_3]$) has not been perturbed in ref.~\cite{xiao-mfu}, it 
can be done with an appropriate parameterization scheme. The Euler angle with
$z$-$x'$-$z''$ convention~\cite{goldstein80euler} is used to parameterize the orientation of the 
Reynolds stress tensor. That is, first, the local coordinate system $x$-$y$-$z$ of
eigenvectors of $[\mathbf{R}]$ initially aligned with the global coordinate system $X$-$Y$-$Z$
rotates about the $z$ axis by angle $\varphi_1$. Then, it rotates about the $x$ axis by 
angle $\varphi_2$, and finally rotates about its new $z$ axis by angle $\varphi_3$. 

In summary, the Reynolds stress tensor field is
transformed to six physically meaningful component fields denoted as $\xi, \eta, k, \varphi_1, \varphi_2$, 
and $\varphi_3$, which are all scalar fields. After the mapping, uncertainties are
introduced to these quantities by adding discrepancy terms to the
corresponding RANS predictions, i.e.,
\begin{subequations}
    \label{eq:delta}
  \begin{alignat}{2}
    \xi (x) & = &\ \tilde{\xi}^{rans}(x) & + \Delta \xi(x)  \\
    \eta(x) & = &\ \tilde{\eta}^{rans}(x) & + \Delta \eta(x)\\
    \log k(x) & = &\ \log \tilde{k}^{rans}(x)  & + \Delta \log k(x)\\
    \varphi_{i}(x) & = &\ \tilde{\varphi_i}^{rans}(x) & + \Delta \varphi_i(x), \ \  i = 1, 2, 3  
    \label{eq:kdelta}    
  \end{alignat}
\end{subequations}
where $\Delta \xi(x)$ and $\Delta \eta(x)$ are discrepancies of the Reynolds stress shape 
parameters $\xi$ and $\eta$, respectively; $\Delta \log k(x)$ is the log-discrepancy of the 
turbulent kinetic energy; $\Delta \varphi_i(x)$ $(i = 1, 2, 3)$ are the discrepancies of
three Euler angles. To model the prior of these discrepancy fields with spatial 
smoothness, we assume that each discrepancy field is normally distributed at any location $x$
and use a Gaussian kernel $K(x, x')$ to describe the correlation between any two different 
locations $x$ and $x'$. That is,
\begin{equation}
  \label{eq:gp-kernel}
  K(x, x') = \sigma(x)  \sigma(x') 
  \exp \left( - \frac{|x - x'|^2}{l^2}  \right) \quad \textrm{where} \; x, x' \in \Sigma, 
\end{equation}
where $\Sigma$ is the spatial domain of the flow field.  The variance $\sigma(x)$ is a spatially
varying field representing the magnitude of injected uncertainties.  The correlation length $l$ also
varies spatially based on the local length scale of the mean flow. It can be seen from
Eq.~\ref{eq:delta} that the discrepancy fields for shape parameters ($\xi$, $\eta$) and Euler angles
($\varphi_i$) are Gaussian random fields, and the discrepancy field for magnitude parameter $k$ is a
log-normal random field. To reduce dimensions of the random fields, Karhunen--Loeve expansions of
the random fields are adopted with chosen basis that are eigenfunctions of the
kernel~\cite{le2010spectral}.  That is, the discrepancies can be represented as follows:
\begin{equation}
  \label{eq:delta-proj}
  \Delta(x) = \sum_{j=1}^\infty \omega_{\alpha} \; \phi_{\alpha} (x) , 
\end{equation}
where the coefficients $\omega_{\alpha}$ (denoting $\omega^{\xi}_{\alpha}$,
$\omega^{\eta}_{\alpha}$, $\omega^k_{\alpha}$, $\omega^{\varphi_i}_{\alpha}$ for discrepancy fields $\Delta \xi$, $\Delta \eta$,
$\Delta \log k$ and $\Delta \varphi_i$, respectively) are independent standard Gaussian random variables. In practice, the
infinite series are truncated to $N_{kl}$ terms with $N_{kl}$ depending on the smoothness of the kernel $K$.
Therefore, the discrepancy fields in Reynolds stress are parameterized by the coefficients
$\omega^\xi_{\alpha},\, \omega^\eta_{\alpha}, \, \omega^k_{\alpha}, \, \omega^{\varphi_i}_{\alpha}$ with $\alpha = 1, 2, \cdots, N_{kl}$. 
These Reynolds stress discrepancies are then propagated to the QoI (e.g., velocity) as the model-form uncertainties. 
 
In summary, the uncertainties are injected into the six physically meaningful dimensions
of Reynolds stress separately in the physics-based approach. For each dimension, the uncertainties
are represented by the Gaussian random fields to ensure the smoothness of the Reynolds stress perturbations 
(for incompressible flow). The perturbed 
Reynolds stresses are reconstructed with these six random fields, which are then propagated to the 
QoI as the quantified model-form uncertainties. Detailed algorithm of the physics-based approach is presented in \ref{apped:phy}.

\subsection{Random Matrix Theoretic Approach with Maximum Entropy Principle}
\label{sec:RMT}
Since Reynolds stresses belong to the set $\mathbb{M}_d^{+0}$ of symmetric positive definite
tensors, where $d = 3$, it is natural to describe them directly as random matrices in set
$\mathbb{M}_d^{+0}$. In the following, we summarize the uncertainty quantification approach based on
random matrix theory and maximum entropy principle proposed by Xiao et al.~\cite{xiao-mfu4}. In this
framework, a probabilistic model in matrix set is built to satisfy all the constraints in the
context of turbulence modeling. Based on the principle of maximum entropy, the target probability
measure $p_{[\text{\textbf{\tiny R}}]} : \mathbb{M}_d^{+0} \mapsto \mathbb{R}^{+}$ is the most
non-committal probability density function (PDF) satisfying all available constraints (e.g.,
realizability) but without introducing any other unwarranted constraints, where $\mathbb{R}^{+}$ is
the set of positive real number.

A two-step approach is used to find such maximum entropy distribution for random 
Reynolds stress tensors. First, the PDF for a normalized, positive definite random
matrix~$[\mathbf{G}]$ is obtained, whose mean is the identity matrix, i.e., $\mathbb{E}\{ [\mathbf{G}] \} = [I]$. 
The distribution of~$[\mathbf{G}]$ should have the maximum entropy. Second, 
the distribution of Reynolds stress tensor~$[\mathbf{R}]$ is obtained with its mean~$[\underline{R}]$ and 
normalized random matrix~$[\mathbf{G}]$ as follows:
\begin{equation}
  \label{eq:a-transform}
[\mathbf{R}] = [\underline{L}_R]^T  [\mathbf{G}]  [\underline{L}_R] , 
\end{equation}
where $ [\underline{L}_R]$ is an upper triangular matrix with non-negative diagonal entries obtained
from the following factorization of the specified mean~$[\underline{R}]$, i.e.,
\begin{equation}
  \label{eq:lu-decomp}
  [\underline{R}] = [\underline{L}_R]^T [\underline{L}_R].
\end{equation}
It is assumed that the RANS predicated Reynolds stress~$[R]^{rans}$ is the best estimate of~$[\underline{R}]$.
Therefore, to determine the maximum entropy distribution of~$[\mathbf{R}]$, one only needs to focus
on the normalized random matrix~$[\mathbf{G}]$. It is first represented
by its Chelosky factorization:
\begin{equation}
    \label{eq:G}
    [\mathbf{G}] = [\mathbf{L}]^T [\mathbf{L}] ,
  \end{equation}
where $[\mathbf{L}]$ are upper triangle matrices with six independent elements. To achieve the matrix entropy, 
the distributions for each element~$\mathbf{L}_{i j}$ can be specified following the procedures in Ref.~\cite{xiao-mfu4}.  
Note that each of the six elements of~$[\mathbf{L}]$ is a random scalar field independent of each other. 
The off-diagonal element fields~$\mathbf{L}_{i j}(x)$ with $i < j$ are obtained from
    \begin{equation}
      \label{eq:L-offdiag}
      \mathbf{L}_{i j}(x) = \sigma_d(x)  \; \mathbf{w}_{i j}(x), 
    \end{equation}
in which $\mathbf{w}_{i j}(x)$ (i.e., $\mathbf{w}_{12}(x)$, $\mathbf{w}_{13}(x)$, and $\mathbf{w}_{23}(x)$) 
represents an independent Gaussian random field with zero mean and unit variance.
The uncertainty magnitude field~$\sigma_d(x)$ can be calculated as 
\begin{equation}
\label{eq:sigmad}
\sigma_d(x) = \delta(x) \times (d+1)^{-1/2}, 
\end{equation}
in which $d = 3$ denotes the dimension of tensor, 
and $\delta(x)$ denotes dispersion parameter field. The dispersion parameter at location~$x$
indicates the uncertainty of the random matrix at $x$ and is defined as
\begin{equation}
  \label{eq:delta-def}
\delta(x) = \left[ \frac{1}{d} \mathbb{E}\{ \| [G](x) - [I] \|_F^2 \}  \right]^{\frac{1}{2}},
\end{equation}
where $\| \cdot \|_F$ is Frobenius norm, e.g., $\| G\|_F = \sqrt{\operatorname{tr}([G]^T [G])}$.  It
can be seen that $\delta(x)$ is analogous to the variance field~$\sigma(x)$ of a scalar random field shown in
the physics-based approach. 
It has been shown in~\cite{soize2000nonparametric} that $0 < \delta(x) < \sqrt{2}/2$ for $d=3$. 
For the three diagonal element fields, each one of them
is generated as follows:
    \begin{equation}
      \label{eq:L-diag}
      \mathbf{L}_{i i}(x) = \sigma_d(x) \sqrt{2 \mathbf{u}_i(x)} \quad \textrm{ with } 
      \; i = 1, 2, 3,
    \end{equation}
where $\mathbf{u}_i(x)$ is a positive valued gamma random field and $\sigma_d(x)$ is defined
above in Eq.~\ref{eq:sigmad}. 

Since Reynolds stresses are correlated at different spatial locations, one needs to model
the correlation structures in the six fields of elements. It is assumed here that both
the off-diagonal and the square root of diagonal terms have the same spatial correlation 
structures. With Gaussian kernel used in this work, the correlation between two spatial
locations $x$ and $x'$ can be shown as
\begin{subequations}
  \label{eq:Lx}
\begin{align}
  \rho_L \{ \mathbf{L}_{i j}(x) , \mathbf{L}_{i j}(x') \}  &= \exp \left[-\frac{|x-x'|^2}{l^2} \right]
  \quad \textrm{ for } \quad i < j  \\
  \rho_L \{ {\mathbf{L}_{i i}^2(x)} , {\mathbf{L}_{i i}^2(x')} \}  &= \exp \left[ -\frac{|x-x'|^2}{l^2} \right],
\end{align}  
\end{subequations} 
where $l$ is the correlation length scale. Note that repeated index does not imply summation here. 
With the defined correlation model, the three independent Gaussian random fields for the off-diagonal 
terms can be generated by using Monte Carlo sampling. Similar to that in the physics-based approach, 
the Karhunen--Loeve expansions are used here for reducing the dimension. To express the non-Gaussian 
random fields used to obtain the diagonal terms (see Eq.~\ref{eq:L-diag}), the Gamma random variable~$\mathbf{u}_i$ at any spatial location~$x$ 
is expanded by polynomial chaos expansion with Gaussian random variables~\cite{sakamoto2002polynomial}
(see \ref{apped:rmt}).
Finally, the constructed distribution of Reynolds stresses can be propagated by RANS equation to 
the QoI to quantify the model-form uncertainties.    

In summary, the uncertainties are directly injected into the Reynolds stress tensors in the set~$\mathbb{M}_d^{+0}$ of 
positive definite matrices in the RMT approach. The principle of maximum entropy is applied to avoid
introducing artificial constraints. The perturbed Reynolds stresses are then propagated to the QoI as the 
quantified model-form uncertainties. Detailed algorithm of the RMT approach can be found 
in \ref{apped:rmt}.

\section{Numerical Results}
\label{sec:res}
\subsection{Cases Setup}
\begin{figure}[!htbp]
  \centering
  \includegraphics[width=0.65\textwidth]{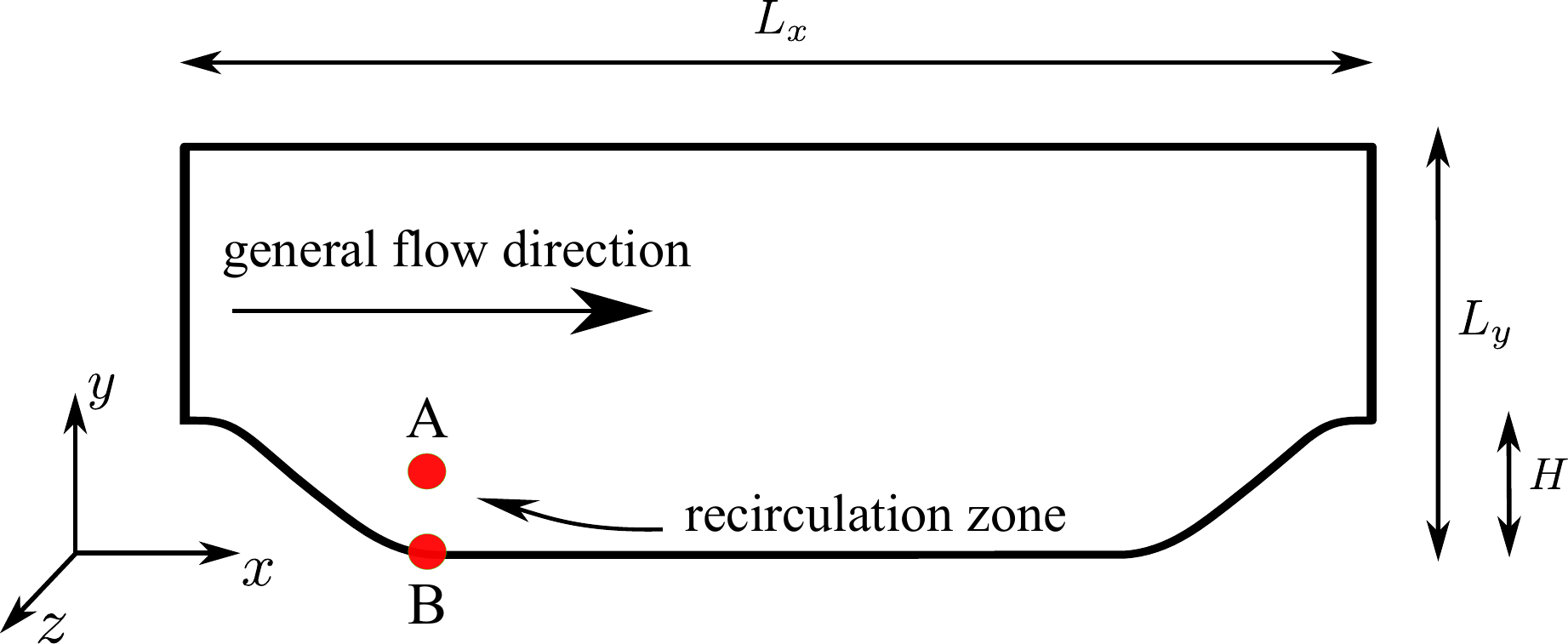}
  \caption{Domain shape for the flow in the channel with periodic hills. The $x$-, $y$- and
    $z$-coordinates are aligned with the streamwise, wall-normal, and spanwise directions,
    respectively. All dimensions are normalized by $H$ with $L_x/H=9$ and $L_y/H=3.036$.
    Two typical locations A  ($x/H = 2.0, y/H = 0.5$)  and B  ($x/H = 2.0, y/H = 0.01$) 
    are marked in the figure.}
  \label{fig:domain_pehill}
\end{figure}

A canonical flow, the flow in a channel with periodic hills, is studied to compare the prior
distributions of Reynolds stresses perturbed by the physics-based approach and 
the random matrix theoretic (RMT) approach. The computational domain is shown in 
Fig.~\ref{fig:domain_pehill}. Periodic boundary conditions are imposed
in the streamwise ($x$-) direction and non-slip boundary conditions are applied at the walls.  

In the physics-based approach the variance fields~$\sigma(x)$ for generating random perturbation
fields $\Delta\xi(x)$, $\Delta\eta(x)$, $\Delta\log k(x)$, and $\Delta \varphi_i(x)$ ($i = 1, 2, 3$)
represent the magnitude of injected uncertainties.  In the RMT approach the dispersion
parameter~$\delta(x)$ determines the variance of perturbations directly in the set~$\mathbb{M}_3^+$
of positive definite matrices.  To ensure that the distributions of perturbed Reynolds stresses from
the two approaches are comparable, the amount of perturbation needs to be consistent with each
other.  Therefore, we estimate the dispersion parameter field~$\delta(x)$ based on the samples of
Reynolds stresses obtained in the physics-based approach with the given
variance~$\sigma(x)$. Although the physics-based approach is more flexible to specify different
variances for the six discrepancy fields, it is difficult to determine how large the variance of
each variable should be relative to each other.  To be consistent with Xiao et al.~\cite{xiao-mfu},
the Gaussian random fields for $\Delta\xi$, $\Delta\eta$, $\Delta\log k$, and $\Delta\varphi_i$ ($i
= 1, 2, 3$) share the same variance field~$\sigma(x)$. Since the aim of this work is to compare the
two approaches, we choose a constant variance field to avoid the complexity caused by spatial
variations of the perturbation variances. We investigate two groups of cases with different
magnitudes of perturbation. For cases Phy1 and RM1, a relatively small constant variance
field~$\sigma = 0.2$ is used in the physics-based approach, and the corresponding dispersion
parameter field~$\delta(x)$ is estimated in the RMT approach.  For cases Phy2 and RM2, a larger
constant variance field~$\sigma = 0.6$ is applied. Since all results below are presented in degrees,
we note that the variances $\sigma = 0.2$ and $\sigma = 0.6$ here for the Euler angles are in
radian, which correspond to $12^\circ$ and $34^\circ$, respectively.  For all cases, 30 modes are
used in Karhunen--Loeve expansion to capture more than $90\%$ of the variance of the Gaussian random
field. To reflect the anisotropy of the flow, an anisotropic yet spatially uniform length scale
($l_x/H = 2$ and $l_y/H =1$) is used in the correlation kernel.  To adequately represent the prior
distribution of Reynolds stresses, 10,000 samples are drawn. The computational parameters are
summarized in Table~\ref{tab:para}.
\begin{table}[htbp]
  \centering
  \caption{Mesh and computational parameters used in the flow over periodic hills.
    \label{tab:para}
  }
     \begin{tabular}[b]{M{5cm}|M{2.1cm} |M{2.4cm} |M{2.1cm} | M{2.4cm}}
      \hline
      \textbf{Parameters} & \textbf{Phy 1} & \textbf{RM 1} & \textbf{Phy 2} & \textbf{RM 2}\\
      \hline
      variance/dispersion$^{(a)}$ & $\sigma = 0.2$ & see note~(a) &  $\sigma = 0.6$ & see note~(a)\\
      \hline
      Karhunen--Loeve mesh & \multicolumn{4}{c}{$50 \times 30$}\\
      RANS mesh & \multicolumn{4}{c}{$50 \times 30$}\\
      number of modes  $N_{\textrm{KL}}$  & \multicolumn{4}{c}{30}\\
      correlation length scales\textsuperscript{(b)}   & \multicolumn{4}{c}{ $l_x/H = 2$, $l_y/H = 1$}\\ 
      number of samples & \multicolumn{4}{c}{10,000} \\
      \hline            
    \end{tabular} \\
  \flushleft
  (a) the dispersion parameter field~$\delta(x)$ is estimated with the corresponding variance field~$\sigma(x)$
  of the companion physics-based case. \\
  (b) see Eq.~\ref{eq:gp-kernel}.
    
\end{table}

\subsection{Results and Discussions}
Two groups of comparisons between the physics-based approach and the 
RMT approach are conducted with the samples drawn by the Monte Carlo method. 
In each case, the samples of Reynolds stress field~$[\mathbf{R}(x)]$
are obtained by using both the physics-based algorithm described in Section~\ref{sec:Phy} 
and the RMT approach in Section~\ref{sec:RMT}. The objective of this study is to gauge the
prior specified in the physics-based approach by using the results of the RMT approach. 
Therefore, all the sampled Reynolds stresses are mapped to their physical dimensions, 
i.e., shape parameter (in Barycentric coordinates $C_1, C_2, C_3$ and natural coordinates $\xi, \eta$),
magnitude~$k$, and the Euler angles ($\varphi_1$, $\varphi_2$, and $\varphi_3$), 
to facilitate the comparison. Moreover, the decomposition into physical components also allows 
visualization of the distribution of random tensor fields. The marginal distributions of all components
are investigated at two typical points: (1) point A located at $x/H = 2.0$ and $y/H = 0.5$ and (2)
point B located at $x/H = 2.0$ and $y/H = 0.01$, indicated in Fig.~\ref{fig:domain_pehill}. 
Point A is a generic point in the recirculation region, and point B is a near-wall point with 
two-dimensional turbulence representing limiting states. 

\begin{figure}[htbp]
  \centering 
   \subfloat[Case Phy1, $\sigma = 0.2$]
  {\includegraphics[width=0.5\textwidth]{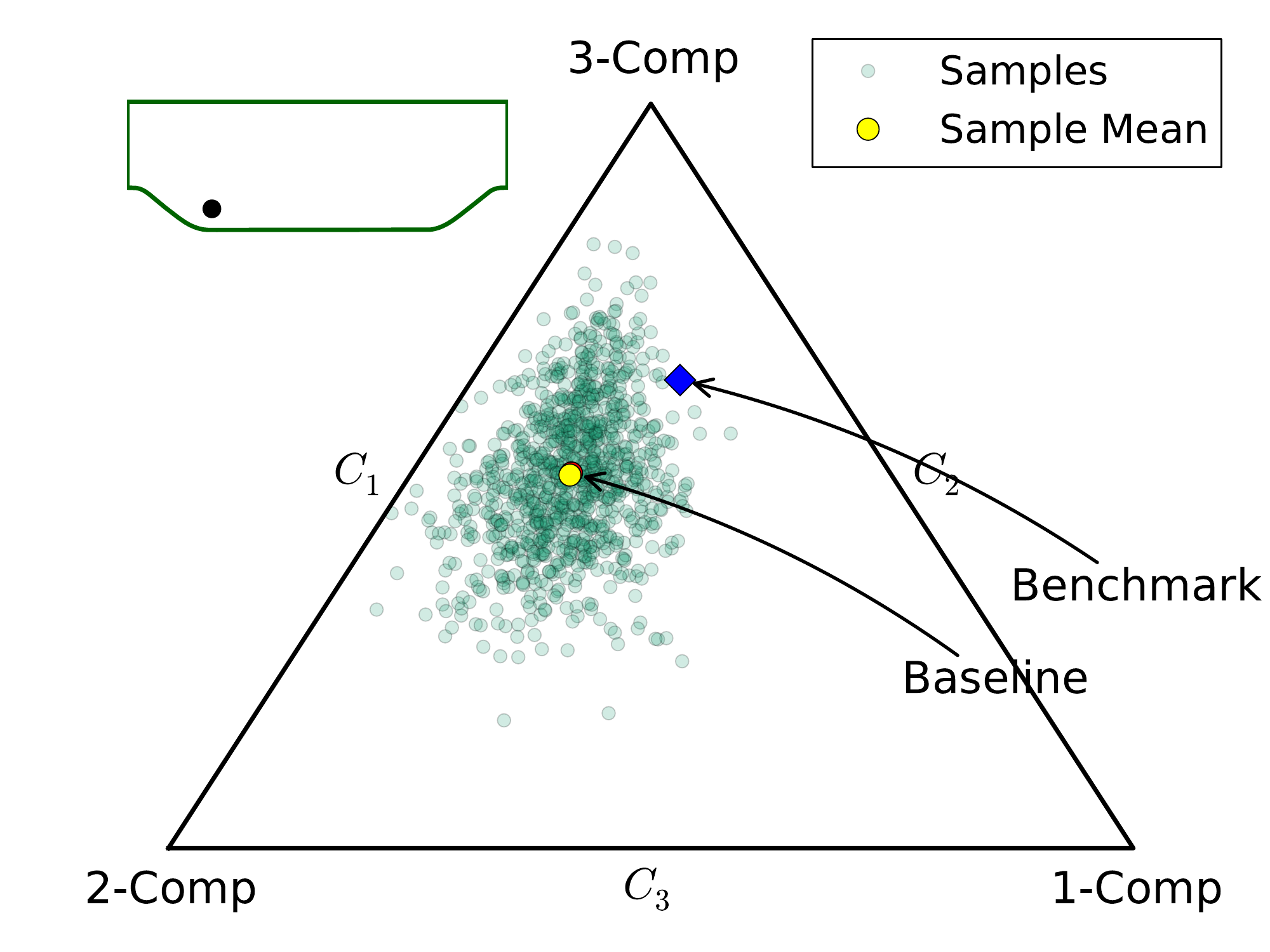}} 
  \subfloat[Case RM1, $\sigma = 0.2$]
  {\includegraphics[width=0.5\textwidth]{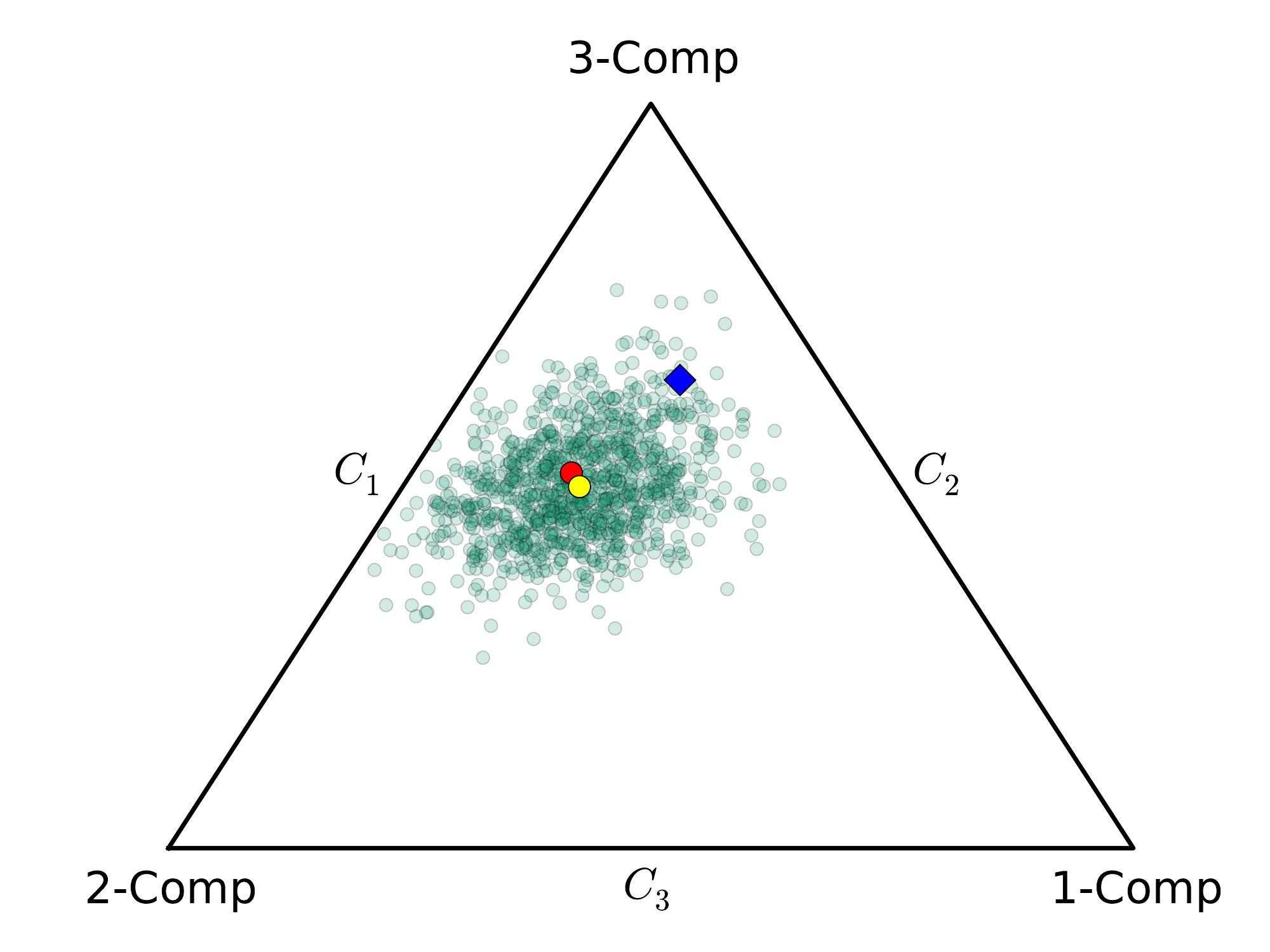}}\\  
  \subfloat[Case Phy2, $\sigma = 0.6$]
  {\includegraphics[width=0.5\textwidth]{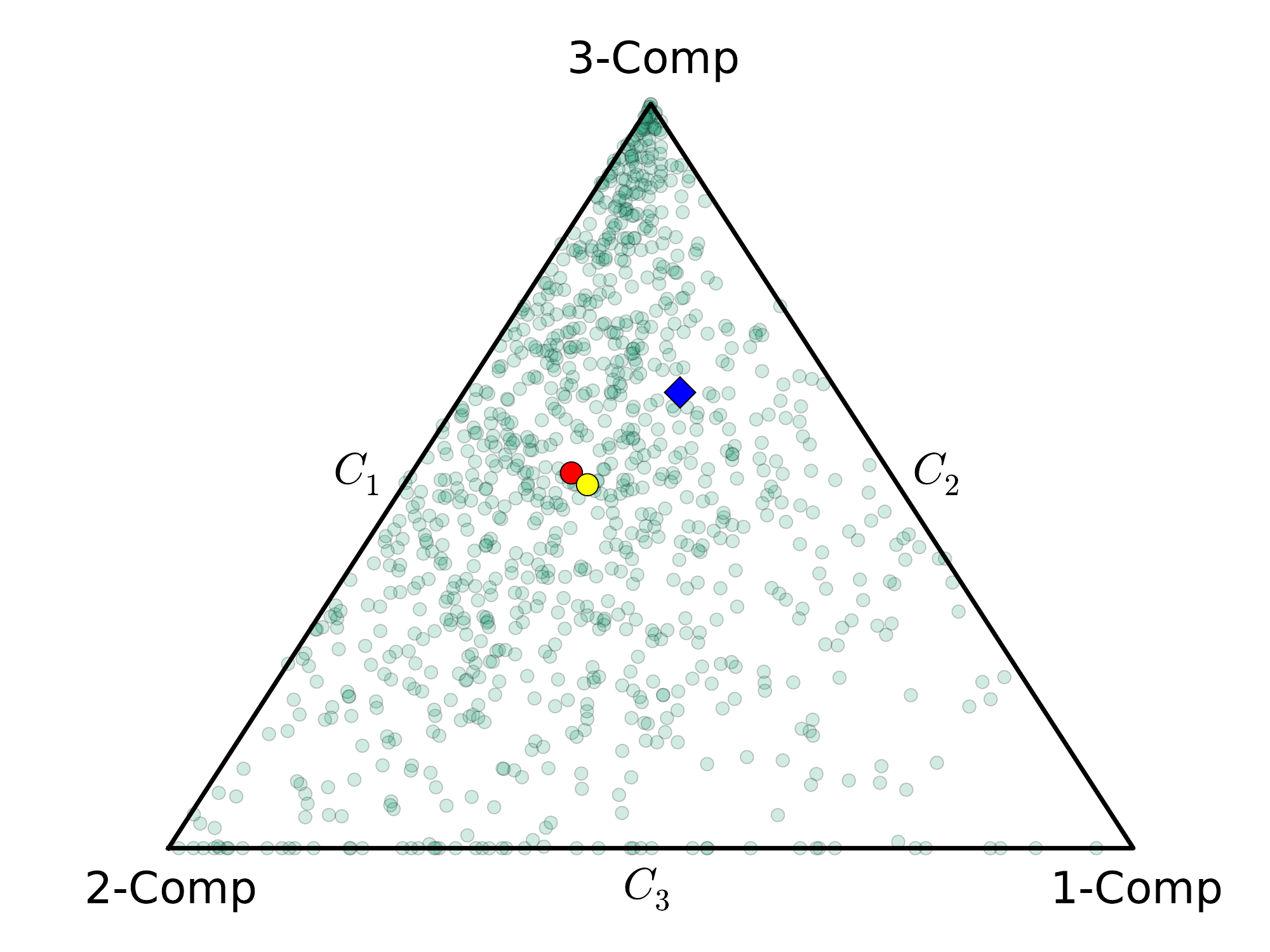}}
  \subfloat[Case RM2, $\sigma = 0.6$]
  {\includegraphics[width=0.5\textwidth]{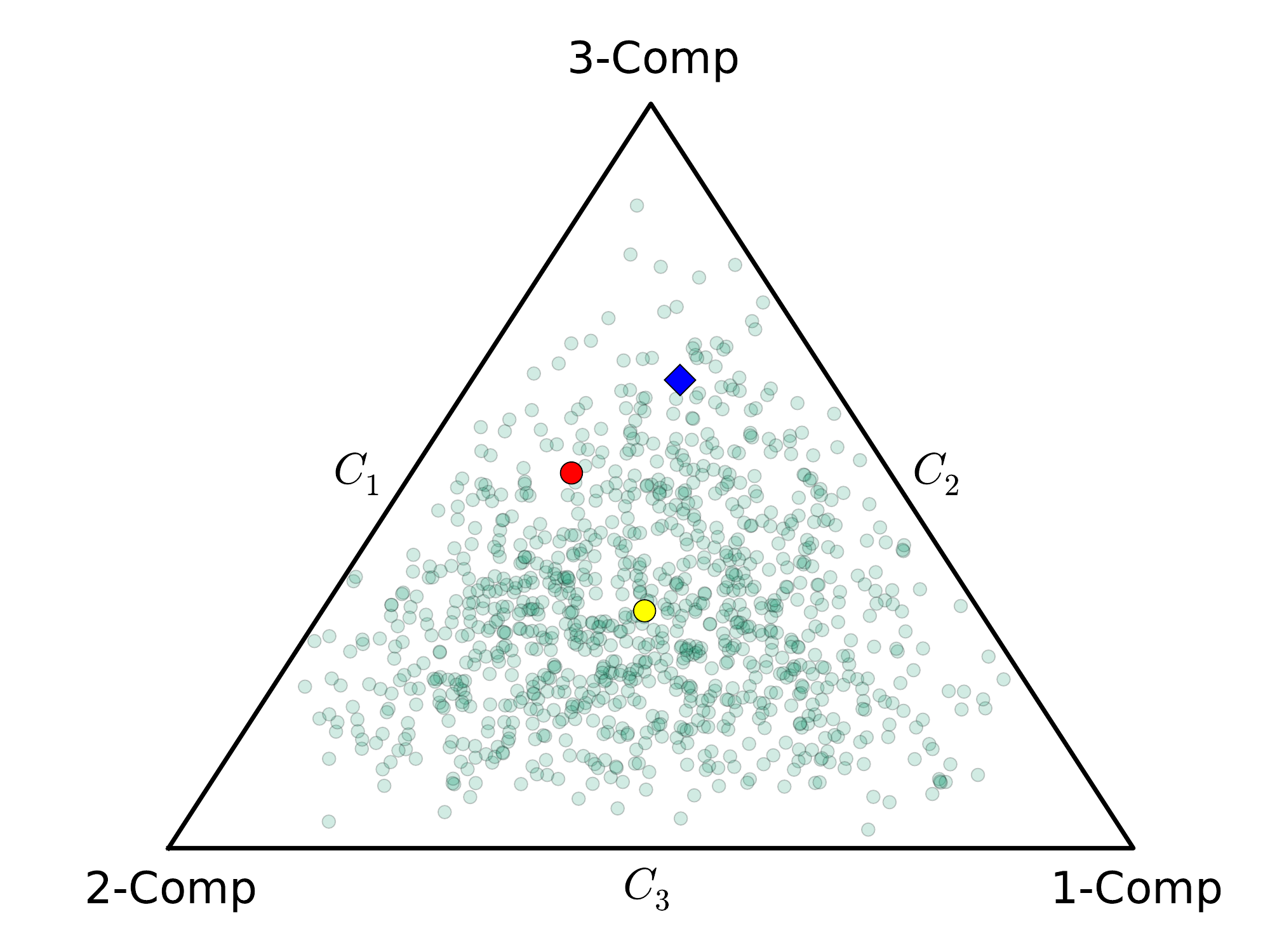}}  
  \caption{ Scatter plots of the Reynolds stress samples projected to 
  the Barycentric coordinates for point A ($x/H = 2.0, y/H = 0.5$) located
  in the recirculation region. Panels (a) and (b) compare
  the two approaches with small perturbation ($\sigma = 0.2$), 
  while panels (c) and (d) show the comparison with large perturbation
  ($\sigma = 0.6$). The benchmark state is plotted as blue square, which is
  located to the upper right of the baseline result.}
  \label{fig:bayLocA}
  \end{figure}

We first investigate the distributions of shape parameters of the Reynolds stress tensors in Barycentric
coordinates, which have clear physical interpretations (see Fig.~\ref{fig:bary}). The perturbed Reynolds 
stresses at the generic point A are sampled and projected onto the Barycentric triangle. The scatter plots
of these samples for all cases are shown in Fig.~\ref{fig:bayLocA}.
We can see that the baseline RANS-predicted Reynolds stress at this generic point 
is located in the interior of the triangle, while the benchmark result is located to 
the upper right of the baseline result. For all cases considered here, all samples fall inside
the Barycentric triangle, demonstrating that the realizability is guaranteed in 
both approaches. The scatterings of samples in Figs.~\ref{fig:bayLocA}a and~\ref{fig:bayLocA}b 
are comparable, and this is also the case for Figs.~\ref{fig:bayLocA}c and~\ref{fig:bayLocA}d, 
indicating that the dispersion parameters~$\delta$ estimated with the 
corresponding variances~$\sigma$ are acceptable. The samples are still scattered around 
the baseline state, especially when the perturbation variance is small ($\sigma = 0.2$).
It shows that the sample mean overlaps with the baseline state in Fig.~\ref{fig:bayLocA}a. 
When the variance increases ($\sigma = 0.6$), the mean value slightly deviates
from the baseline state (comparing Figs.~\ref{fig:bayLocA}a and~\ref{fig:bayLocA}c). 
This deviation is markedly amplified in the RMT approach (Fig.~\ref{fig:bayLocA}d),
suggesting that although the mean of perturbed Reynolds stresses is assumed to be the baseline 
result in the RMT approach, the mean is not preserved during the projection to the Barycentric 
coordinates. As the perturbation becomes large, the constraint of realizability leads to 
this significant deviation. Another notable difference between the results of two approaches 
lies in the shape of sample scattering. The scattering in case Phy1 is more elliptical-like 
with a large amount of samples spreading along the plain strain line (Fig.~\ref{fig:bayLocA}a), 
while in case RM1 the samples are scattered more circularly (Fig.~\ref{fig:bayLocA}b).
When the perturbation is large ($\sigma = 0.6$), we can see that in both cases Phy2 and RM2 
the scatterings of samples show significant influence from the boundaries.  
In the results of the physics-based approach, a number of samples fall on the edges of the triangle, and a large number 
of samples are clustered near the top vertex (Fig.~\ref{fig:bayLocA}c). 
In contrast, in the RMT approach the samples are dispersed within the triangle without 
such clustering, and the frequency of occurrences decreases near the edges  (Fig.~\ref{fig:bayLocA}d).

  \begin{figure}[htbp]
  \centering
  \subfloat[cases Phy1 and RM1, $\sigma = 0.2$]
   {\includegraphics[width=0.5\textwidth]{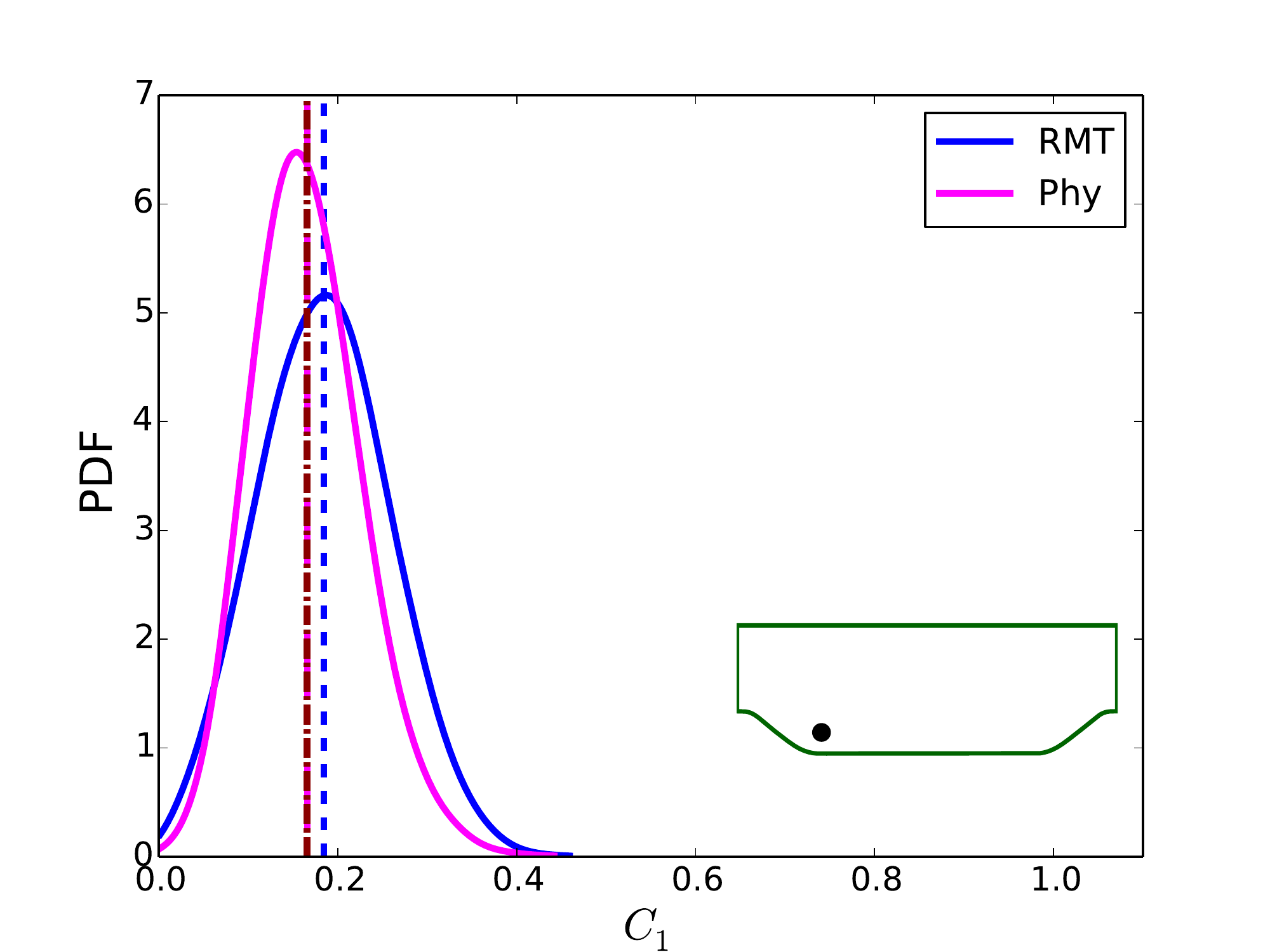}}
   \subfloat[cases Phy2 and RM2, $\sigma = 0.6$]
   {\includegraphics[width=0.5\textwidth]{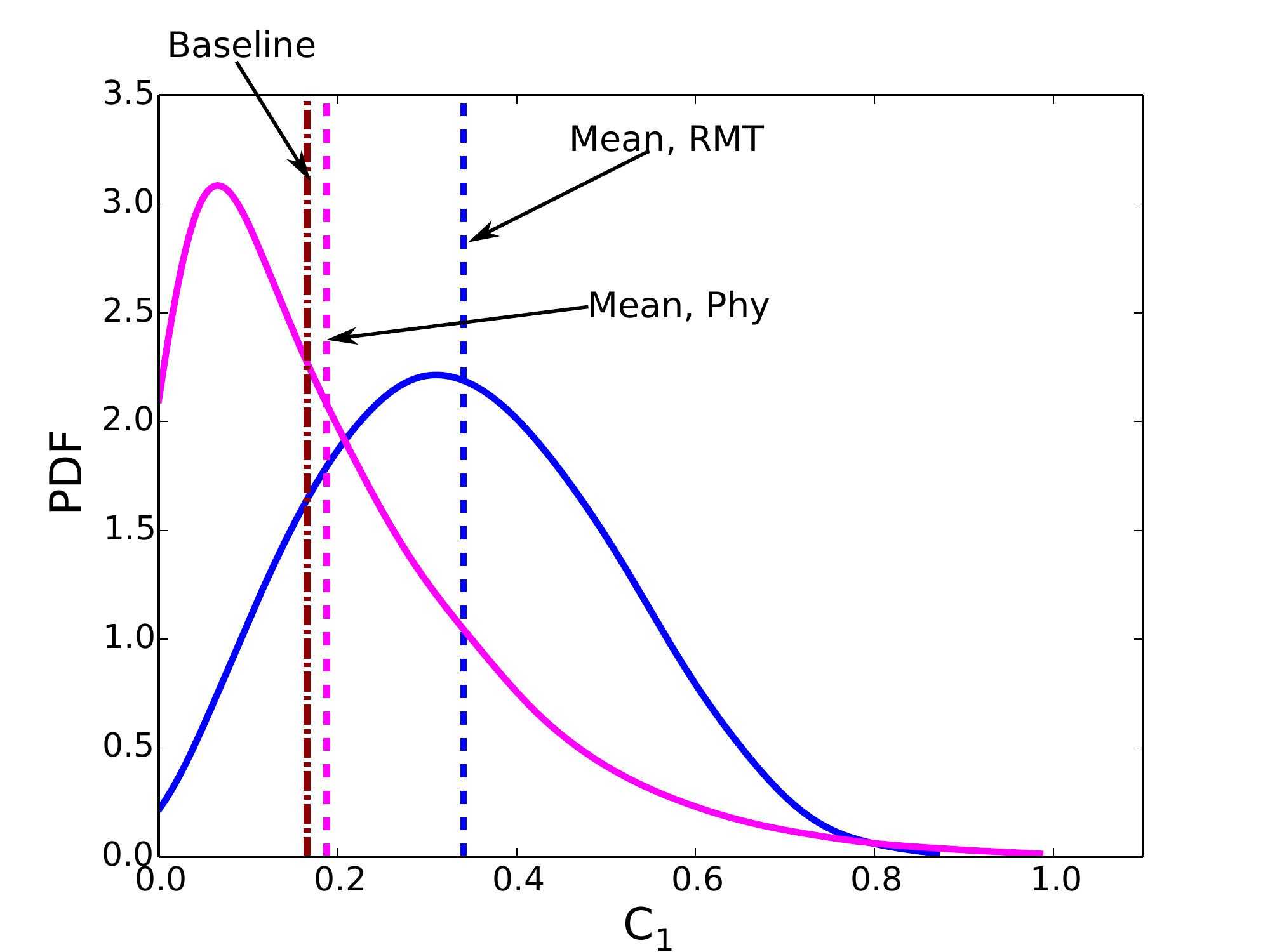}}
   \caption{Probability distributions of the perturbed Barycentric coordinate $C_1$ for point A ($x/H = 2.0, y/H = 0.5$)
   located in the recirculation region. The results from the physics-based approach and the RMT
   approach are compared in the same plot. Panels (a) and (b) show the comparison at perturbation variances
   of $\sigma = 0.2$ (cases Phy1 and RM1) and $\sigma = 0.6$ (cases Phy2 and RM2), respectively. }
   \label{fig:comcon_C1_gen}
  \end{figure} 
 
Detailed comparison can be conducted by examining the marginal distributions of shape parameters 
in Barycentric coordinates $C_1$, $C_2$, and $C_3$. In Fig.~\ref{fig:comcon_C1_gen} we
present the probability density function (PDF) of $C_1$ for cases Phy1 and RM1  and cases Phy2 and RM2. 
Since $C_2$ and $C_3$ are correlated and have the similar 
characteristics as $C_1$, thus are omited for brevity. When the variance~$\sigma$ 
is 0.2, the distributions obtained in both approaches are Gaussian, and the sample 
means are close to the baseline result. As the variance $\sigma$ increases to 0.6, 
both distributions deviate from Gaussian. In the physics-based approach the mode of samples 
(at peak of PDF) moves towards $C_1 = 0$ and the sample mean increases slightly compared to 
the baseline results. This is caused by the scheme used to impose the realizability constraint in the physics-based approach,
where the samples falling outside the triangle are capped to the boundaries of 
natural coordinates (the four edges shown in Fig.~\ref{fig:bary}b). Moreover, 
the samples spreading within the upper area of the natural coordinate square are squeezed in 
Barycentric triangle because of the mapping between the two coordinates.   
However, in the RMT approach the realizability of the Reynolds stress tensor~$[\mathbf{R}]$
is guaranteed mathematically and no additional constraints are imposed due to the coordinates mapping.
More precisely, the positive semi-definiteness of the normalized random matrix~$[\mathbf{G}]$ 
is guaranteed by constructing from its Cholesky factor~$[\mathbf{L}]$ (see Eq.~\ref{eq:G}). 
As a result, the distribution of $C_1$ is close to Gamma distribution with the 
sample mean increased compared to the baseline result.  Based on the observations and discussions above, 
we find the bounding method used and the mapping between natural coordinate and Barycentric coordinate
used in physics-based approach introduce some artificial constraints into the prior of Reynolds stresses. 
Therefore, it is better to specify Gaussian distributed perturbations of shape parameters 
in Barycentric coordinate to achieve maximum entropy for a generic location away from the wall.  
 
\begin{figure}[htbp]
  \centering 
   \subfloat[Case Phy1, $\sigma = 0.2$]
  {\includegraphics[width=0.48\textwidth]{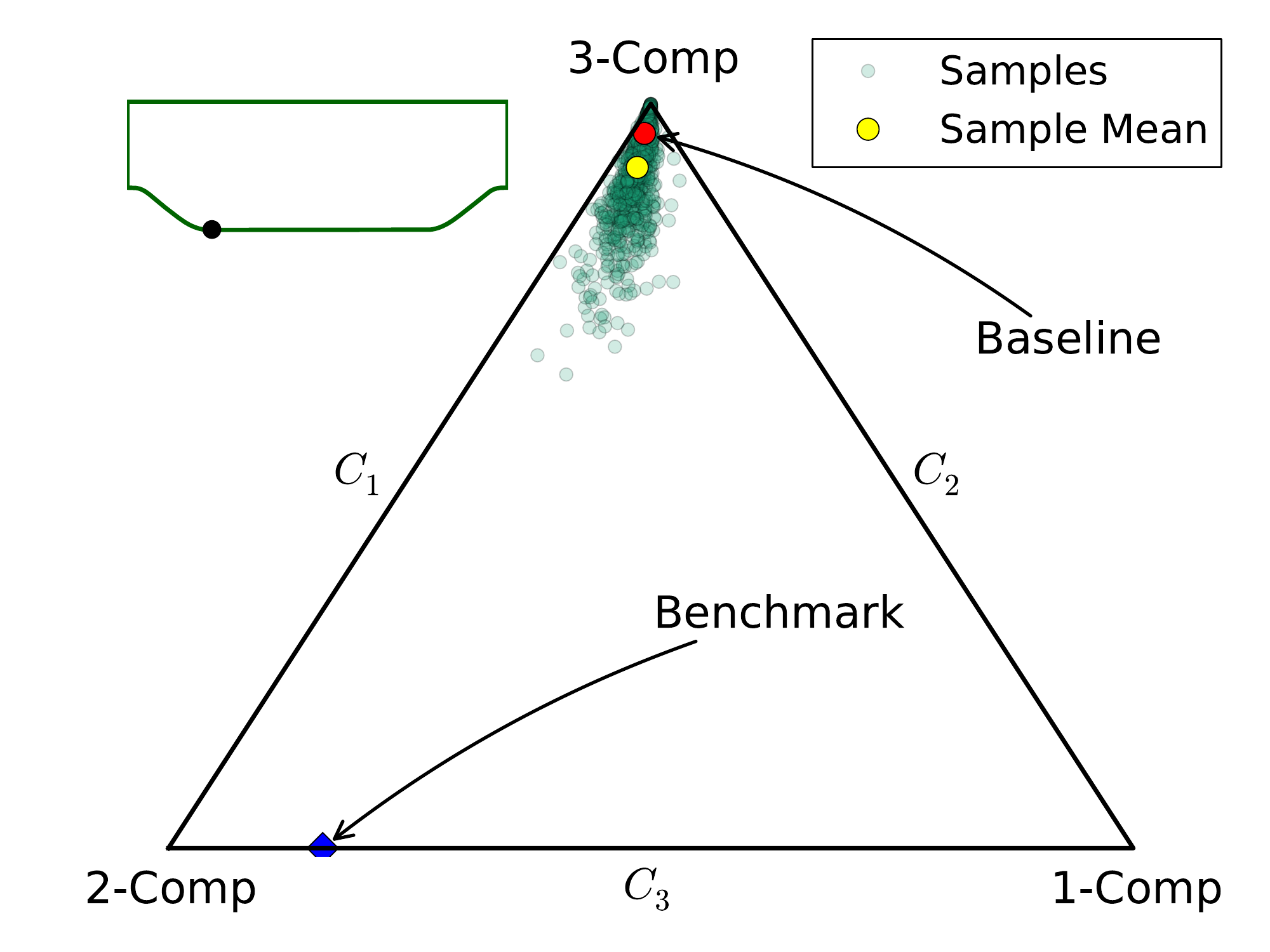}} 
  \subfloat[Case RM1, $\sigma = 0.2$]
  {\includegraphics[width=0.48\textwidth]{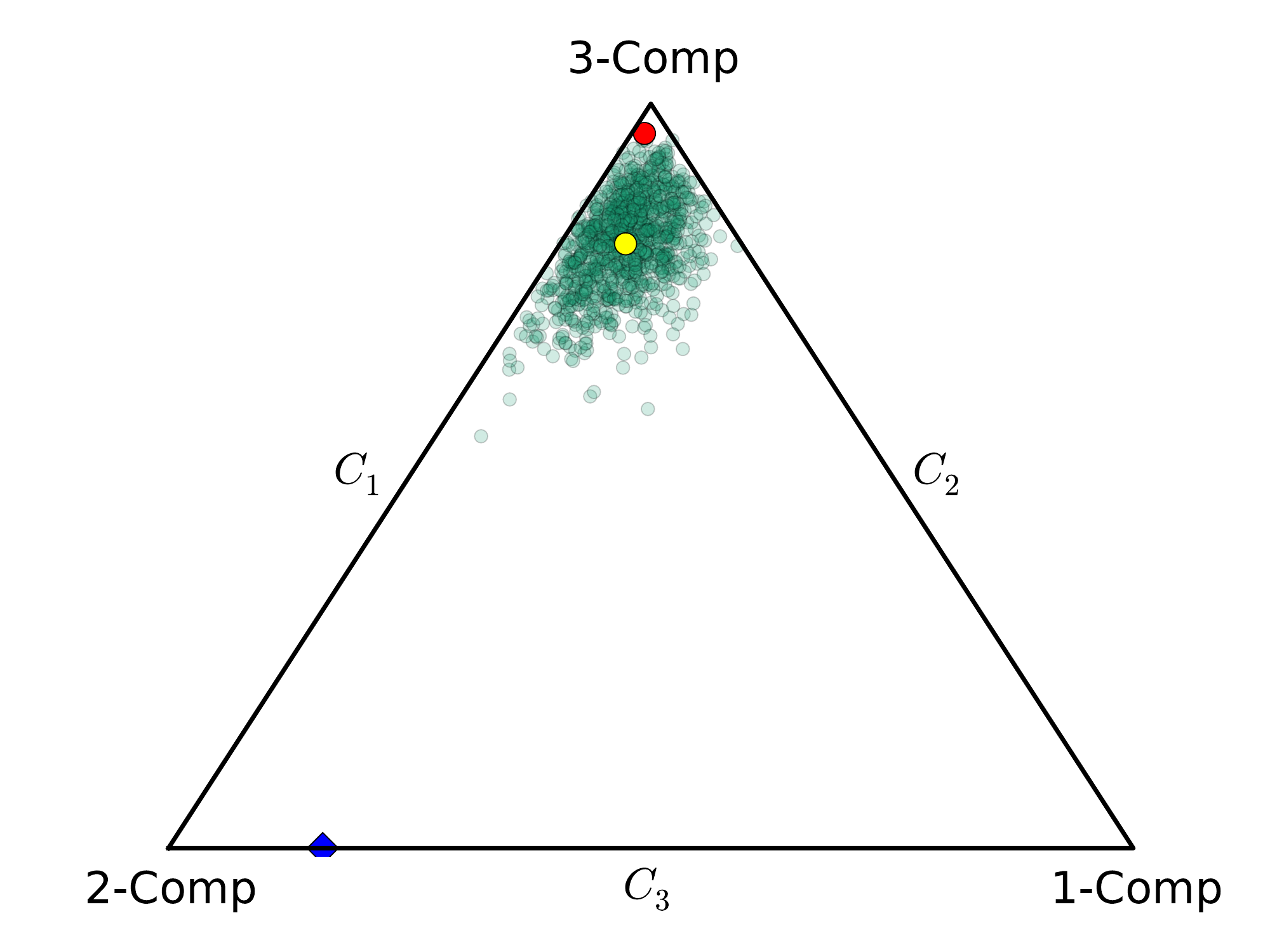}}\\
  
  \subfloat[Case Phy2, $\sigma = 0.6$]
  {\includegraphics[width=0.48\textwidth]{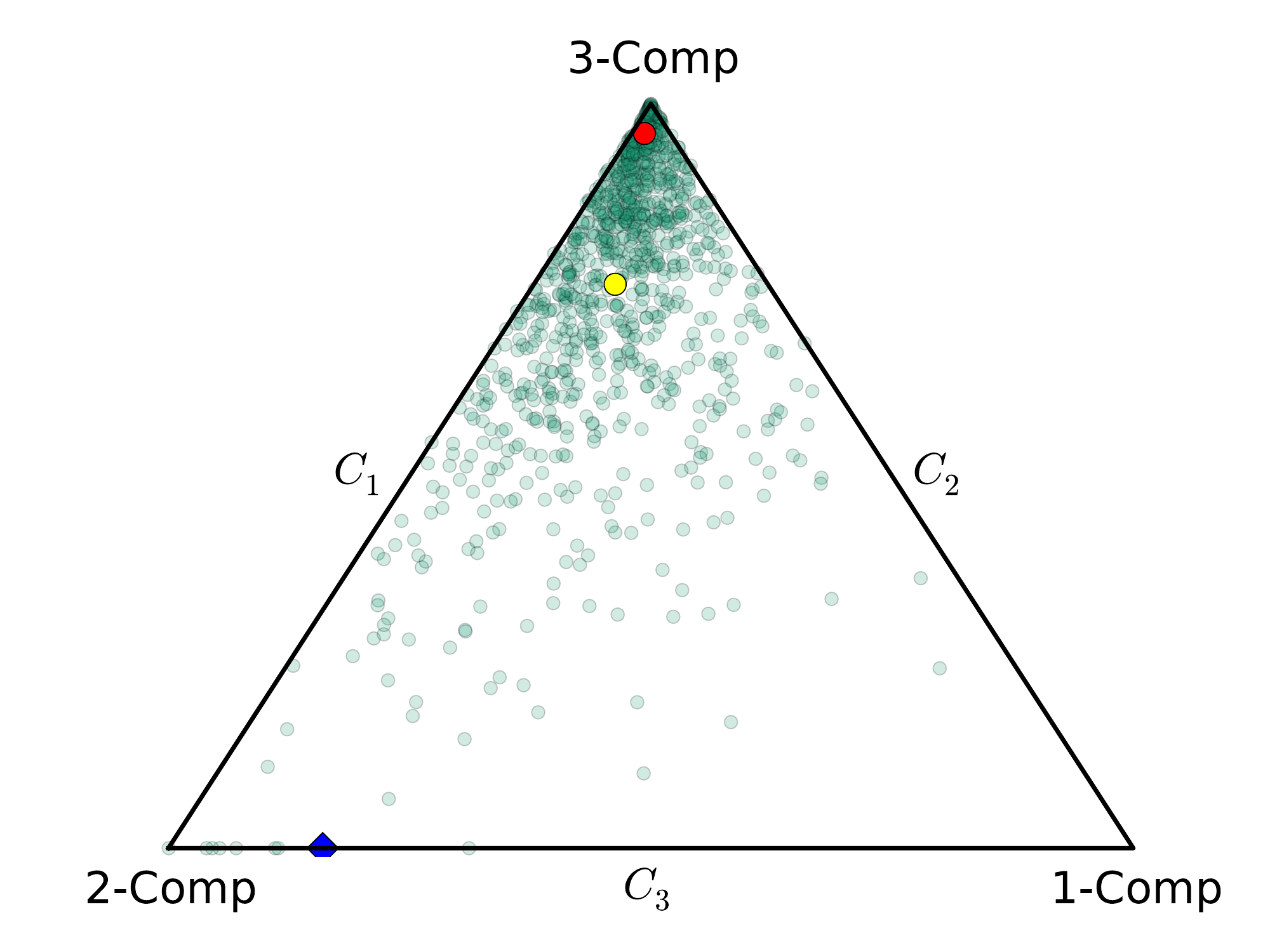}}
  \subfloat[Case RM2, $\sigma = 0.6$]
  {\includegraphics[width=0.48\textwidth]{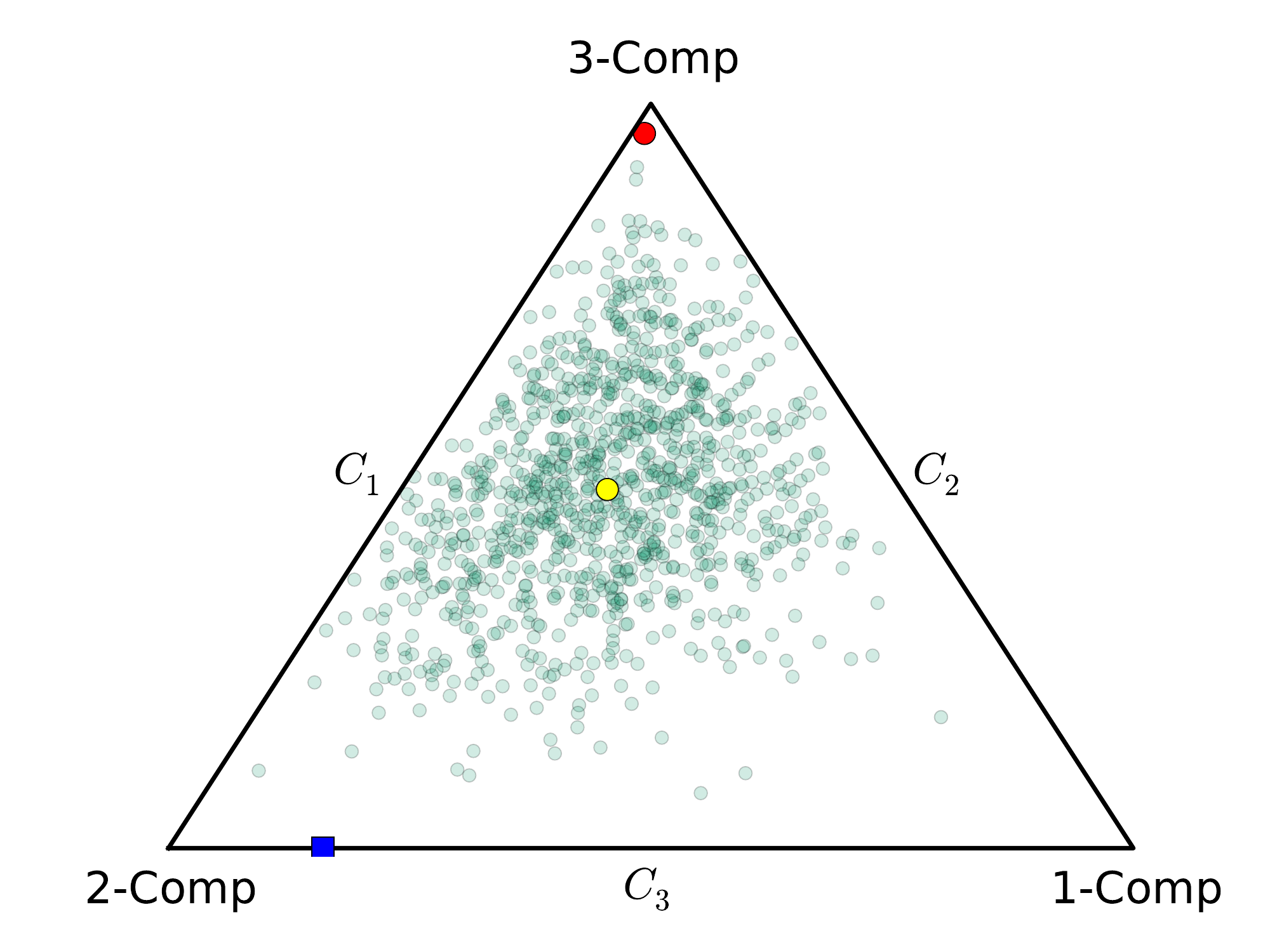}}  
  \caption{ Scatter plots of the Reynolds stress samples projected to 
  the Barycentric coordinates for point B ($x/H = 2.0, y/H = 0.01$) located
  in the recirculation region. Panels (a) and (b) compare
  the two approaches with small perturbation ($\sigma = 0.2$), 
  while panels (c) and (d) show the comparison with large perturbation
  ($\sigma = 0.6$). The benchmark state is plotted as blue square, which is
  located on the bottom edge of the triangle, indicating two-component turbulence.
}
  \label{fig:bayLocB}
  \end{figure}

It is also interesting to study the a point located close to the wall. 
Similarly, the comparisons of scatter plots of Reynolds stress samples at point B are presented 
in Fig.~\ref{fig:bayLocB}. It shows that the benchmark truth is located on the 
bottom edge of the triangle, indicating the two-component limiting state.  
This is because the velocity fluctuations in wall-normal direction are suppressed by
the blocking of bottom wall of the channel. In contrast, RANS-predicted Reynolds stress 
is close to the three-component isotropic state, located near the top vertex of the triangle. 
For the point B at the near wall location, the sample scatterings in all four cases are markedly affected
by the boundaries of the triangle, and the sample means move downwards. 
This is due to the fact that the distances from the 
baseline state to the boundaries are relatively small compared to the perturbations, 
and thus the perturbed states are significantly affected by these constraints. 
However, the influences caused by the constraints imposed in the two approaches 
are different.  In the case with a small variance~$\sigma = 0.2$~(case Phy1, Fig.~\ref{fig:bayLocB}a), 
the samples are largely clustered near the top
vertex and the scattering is squeezed artificially in the physics-based approach. Although
enough samples are drawn, very few of them fall in the areas near the two side edges. 
In contrast, the samples in the RMT approach are dispersed within
the entire upper area of the triangle and better explored the spanned uncertainty 
space~(case RM1, Fig.~\ref{fig:bayLocB}b). When the variance is large 
($\sigma =0.6$, case Phy2, Fig.~\ref{fig:bayLocB}c), the capping scheme
used to ensure realizability in the physics-based approach has a more
pronounced effect on the obtained sample distribution. Specifically, 
about one half of the samples are capped to the edges or the vertices of
the Barycentric triangle, leading to deteriorated sample effectiveness.  
    
\begin{figure}[htbp]
  \centering
   \subfloat[$\sigma = 0.2$]
   {\includegraphics[width=0.9\textwidth]{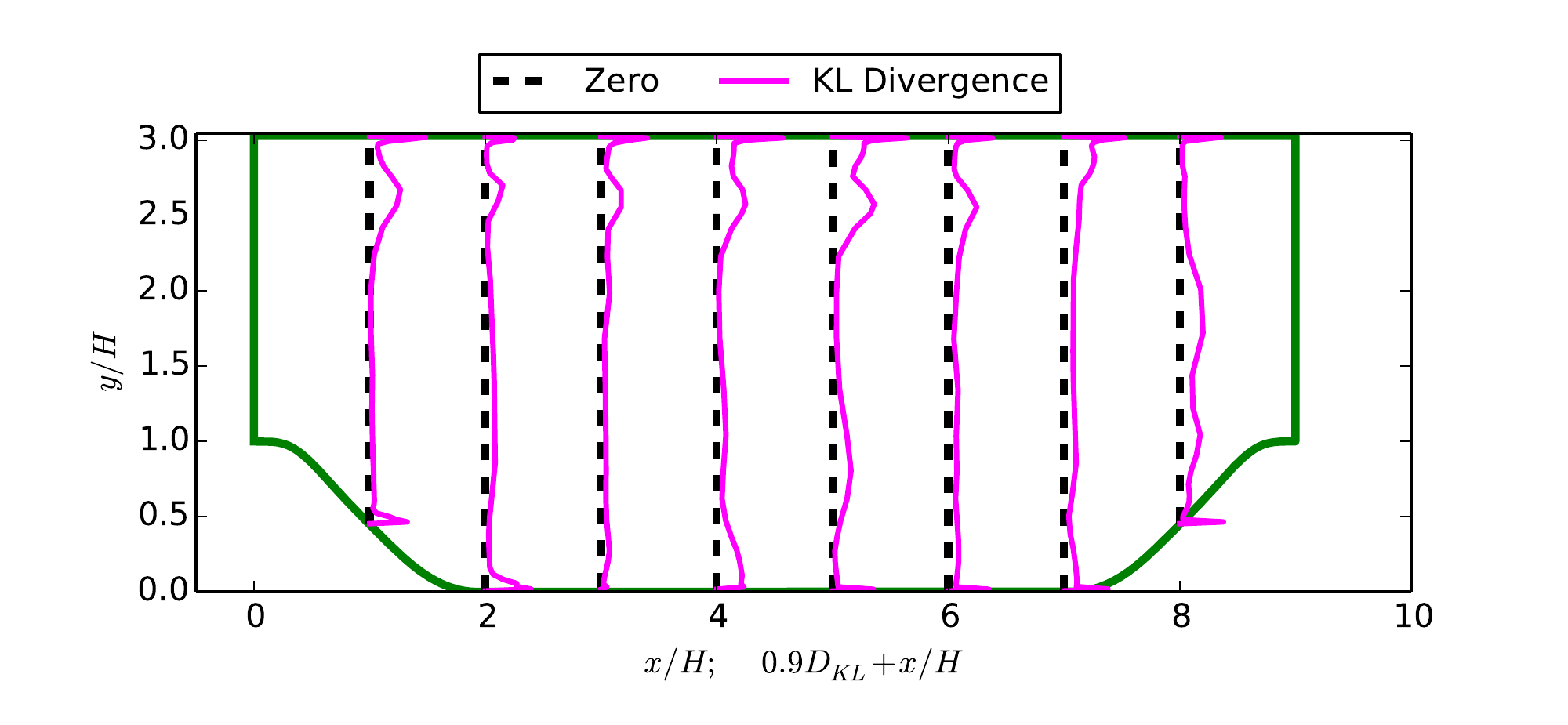}}\\
   \subfloat[$\sigma = 0.6$]
   {\includegraphics[width=0.9\textwidth]{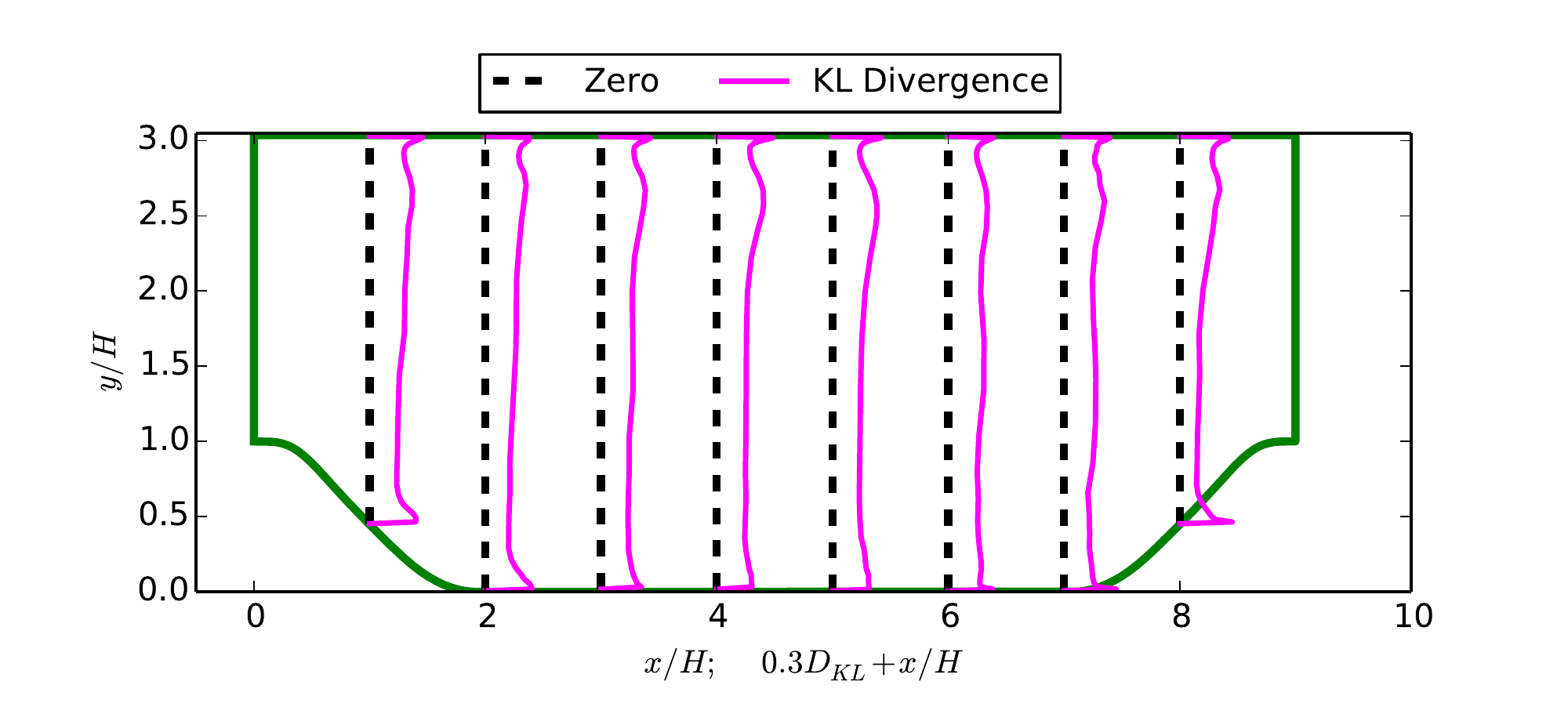}}    
   \caption{Kullbck-Leibler divergence profiles with (a) $\sigma = 0.2$ and (b) $\sigma = 0.6$. 
   The profiles are shown at eight streamwise locations $x/H = 1, \cdots, 8,$ and the reference lines, 
   $D_{KL} = 0 + x/H$, are also plotted. Note that in panel (a), since the scale of $D_{KL}$ is quite smaller
   than that of panel (b), a larger scale factor of 0.9 is used in (a), while a smaller scale factor of 0.3 is used in (b). }
   \label{fig:KL_C1}
  \end{figure}

The differences of perturbations in shape parameters between the two approaches are discussed above
for two typical points A and B of the channel. In order to quantitatively explore the spatial 
variation of the difference between the two approaches, we calculate the Kullbck-Leibler
divergence of the distribution obtained by the physics-based approach from the distribution
obtained by the RMT approach. The Kullbck-Leibler divergence (also known as relative entropy) 
of $q$ from $p$ is a measure of the difference between two probability densities $p$ and $q$, 
which is defined as~\cite{kullback1968information}
\begin{equation}
D_{KL}(p||q) = \int_{\mathbb{I}}\log\frac{p(\vartheta)}{q(\vartheta)}d\vartheta, 
\end{equation}
where $\vartheta$ denotes the parameter, and $\mathbb{I}$ is the parameter space. The Kullbck-Leibler divergence
is analogous to a ``distance" between two distributions, but it is not a distance measure since it does not 
preserve the symmetry in $p$ and $q$. More intuitively, the Kullbck-Leibler divergence of $q$ from $p$ can be interpreted as 
the measure of the information gained from the distribution $p$ to distribution $q$. The Kullbck-Leibler divergence 
is calculated to reflect the additional information introduced in the physics-based approach
based on the maximum entropy distribution obtained with the RMT approach. 
Figure~\ref{fig:KL_C1} shows the spatial profiles of Kullbck-Leibler divergence for the shape parameter~$C_1$. 
The profiles are shown at eight streamwise locations, $x/H = 1, \cdots, 8$, and the dashed black 
lines are plotted to indicate the axis of~$D_{KL} = 0$. 
The geometry of the physical domain is also plotted to facilitate visualization. 
To clearly show the characteristics of the profiles, 
a larger scale factor of 0.9 is used in Fig.~\ref{fig:KL_C1}a, while a smaller scale factor of 0.3 is 
used in Fig.~\ref{fig:KL_C1}b. 
When the variance is small ($\sigma = 0.2$), the Kullbck-Leibler divergence is also small over the entire
domain (Fig.~\ref{fig:KL_C1}a), suggesting that the distributions of shape parameters obtained 
from the physics-based approach are similar to those obtained from the RMT approach. 
When we increase the perturbation variance ($\sigma = 0.6$), the Kullbck-Leibler divergence becomes
larger over the entire domain (Fig.~\ref{fig:KL_C1}b), indicating that the additional, artificial constraints
introduced in the physics-based approach are significant with large perturbation. Moreover, 
both Figs.~\ref{fig:KL_C1}a and~\ref{fig:KL_C1}b show 
that the Kullbck-Leibler divergences for the locations near the wall are
slightly larger than those at generic locations. As has been discussed above, the baseline
RANS-predicted Reynolds stress near the wall is close to three-component isotropic limiting state. This
results in the fact that the perturbation is large compared to the distance from baseline state to the 
boundary of the triangle, and thus more artificial information is introduced in the physics-based
approach at the near-wall regions. It is noted that the $D_{KL}$ is also large at $y/H = 2.5$, 
which is because the baseline state is closer to the top vertex of triangle
at $y/H = 2.5$. All these observations are consistent with the discussion above 
for the two typical points A and B. 
 
 \begin{figure}[htbp]
  \centering
   \subfloat[Phy1 and RM1 ($\sigma = 0.2$)]
   {\includegraphics[width=0.5\textwidth]{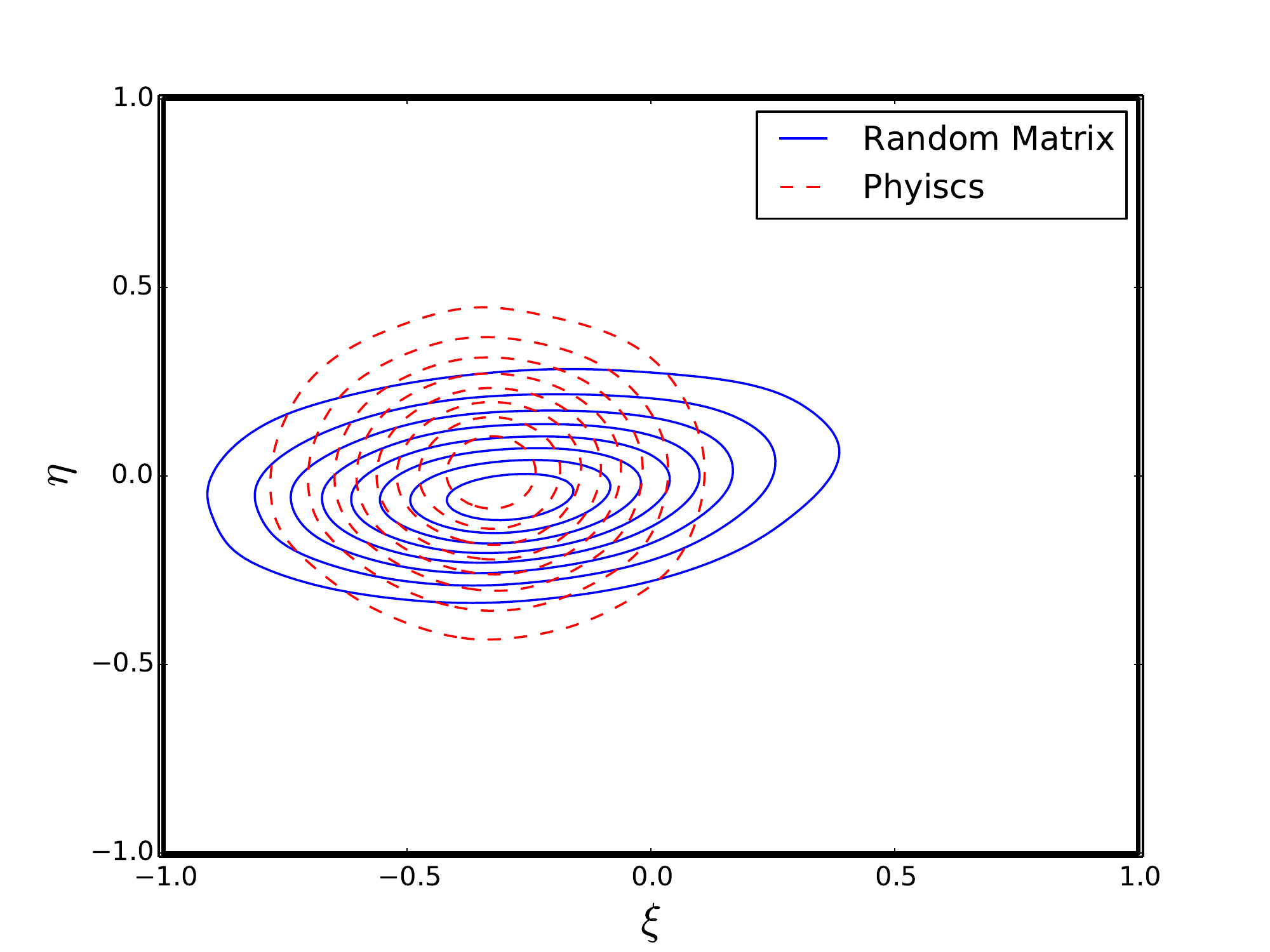}}
   \subfloat[Phy2 and RM2 ($\sigma = 0.6$)]
   {\includegraphics[width=0.5\textwidth]{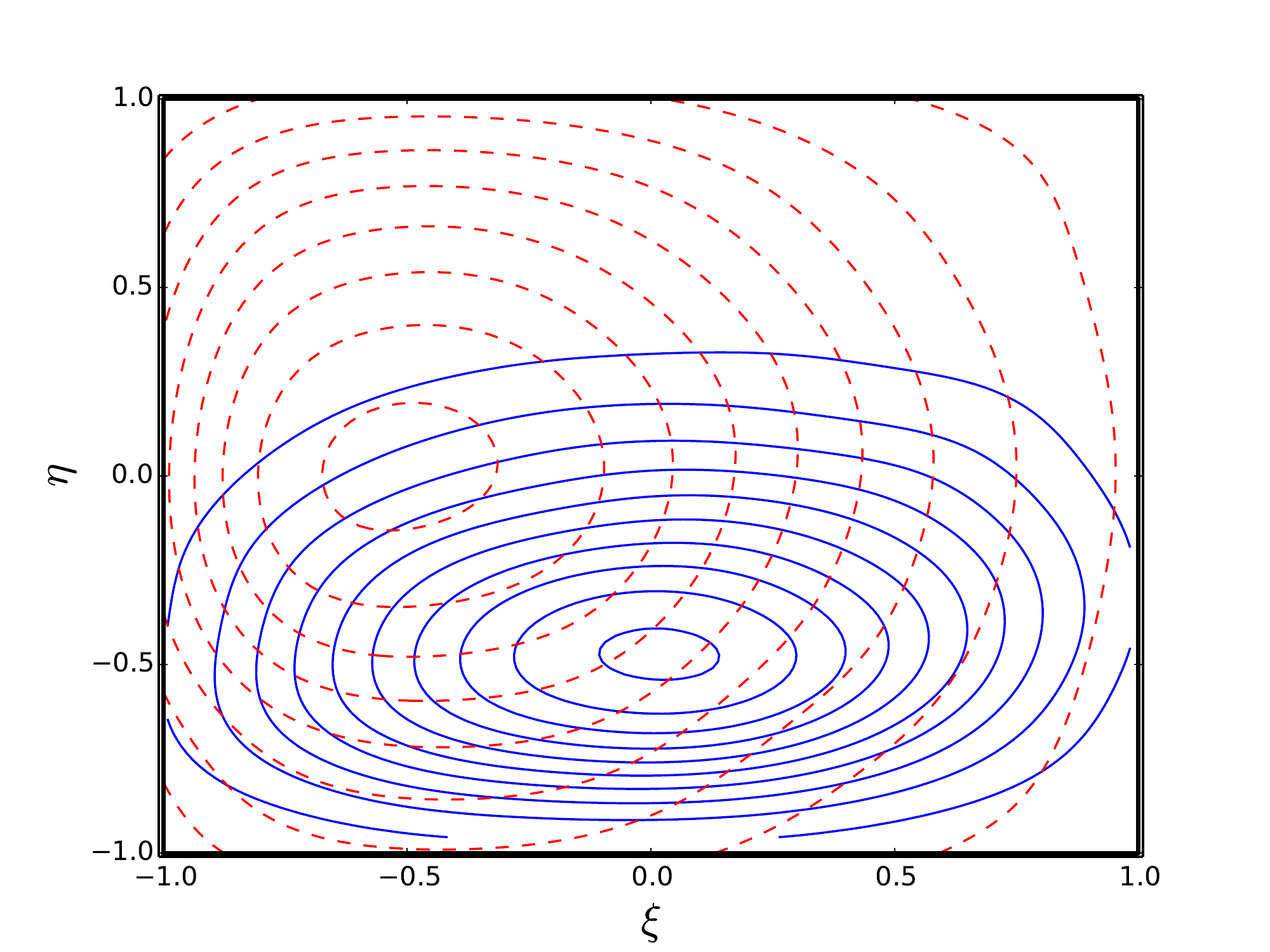}}    
   \caption{Comparisons of probability density contours of natural coordinates ($\xi$ and $\eta$) obtained 
   from the physics-based approach and the RMT approach at point A ($x/H = 2.0, y/H = 0.01$.
   Panel (a) and (b) are with the perturbations $\sigma = 0.2$, and $\sigma = 0.6$, respectively.)}
   \label{fig:contourNat_gen}
  \end{figure} 
  
  \begin{figure}[htbp]
  \centering
  \subfloat[PDF  of $\xi$, $\sigma = 0.2$]
   {\includegraphics[width=0.5\textwidth]{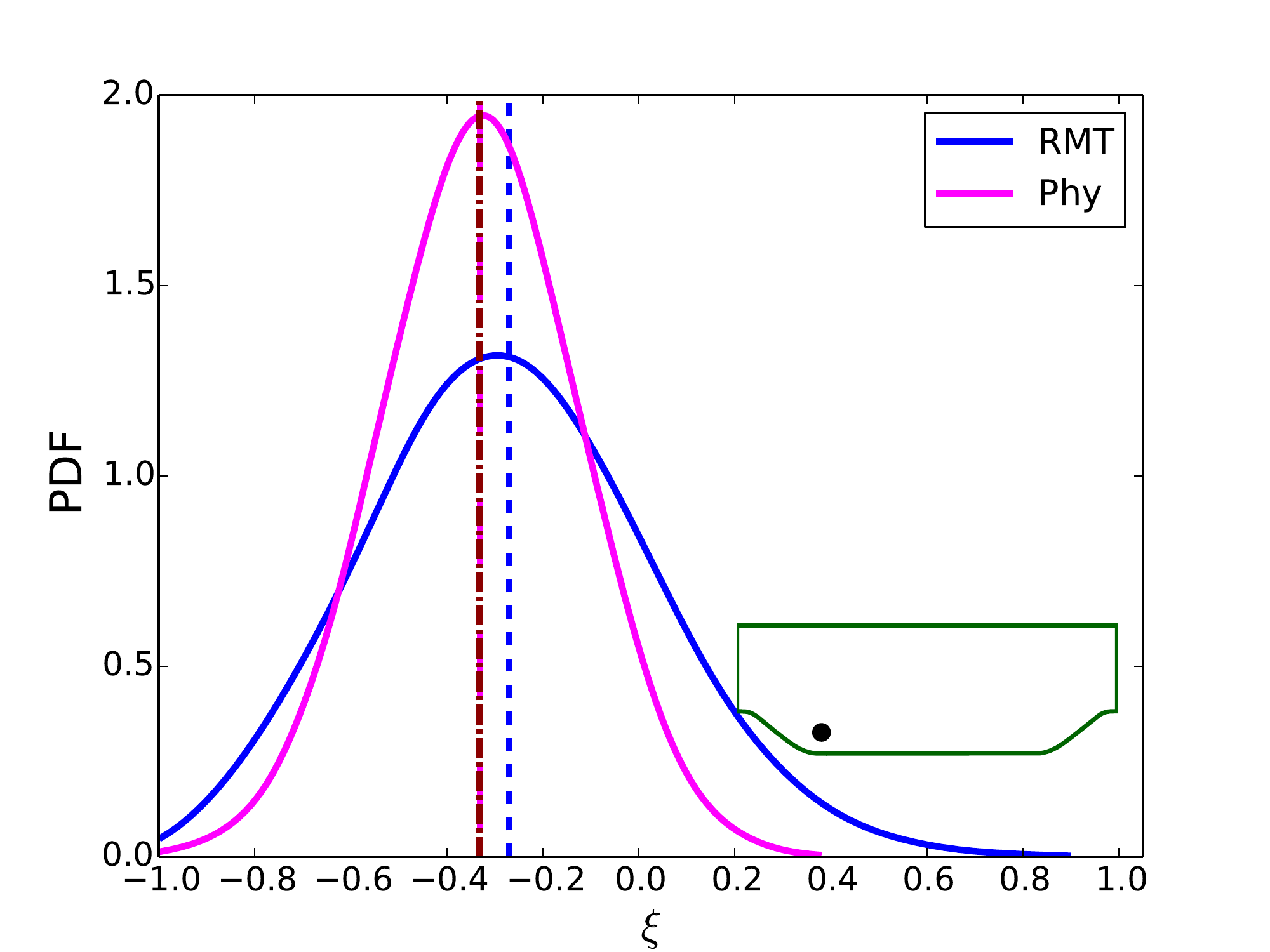}}
   \subfloat[PDF  of $\xi$, $\sigma = 0.6$]
   {\includegraphics[width=0.5\textwidth]{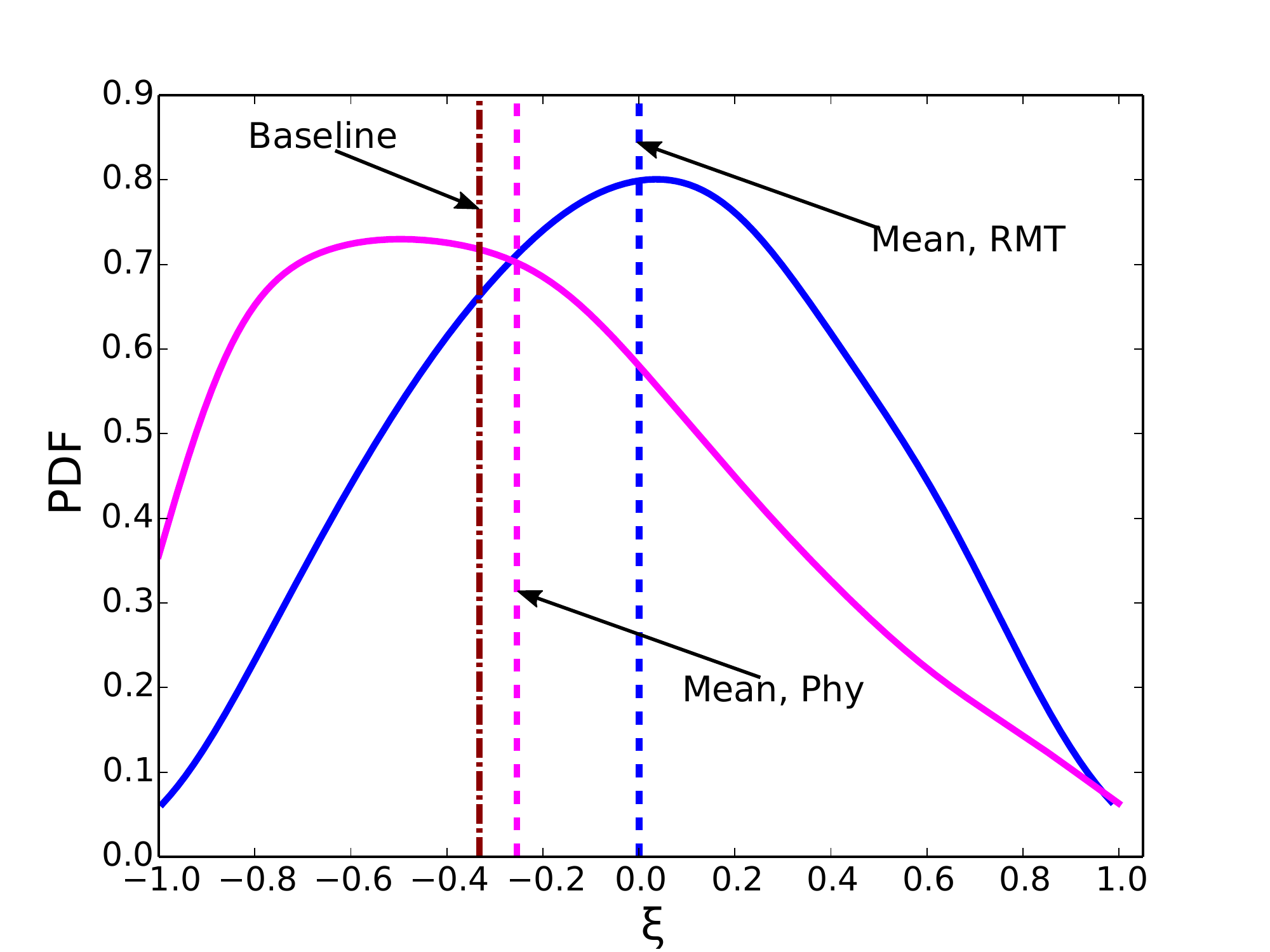}}\\
  \subfloat[PDF  of $\eta$, $\sigma = 0.2$]
   {\includegraphics[width=0.5\textwidth]{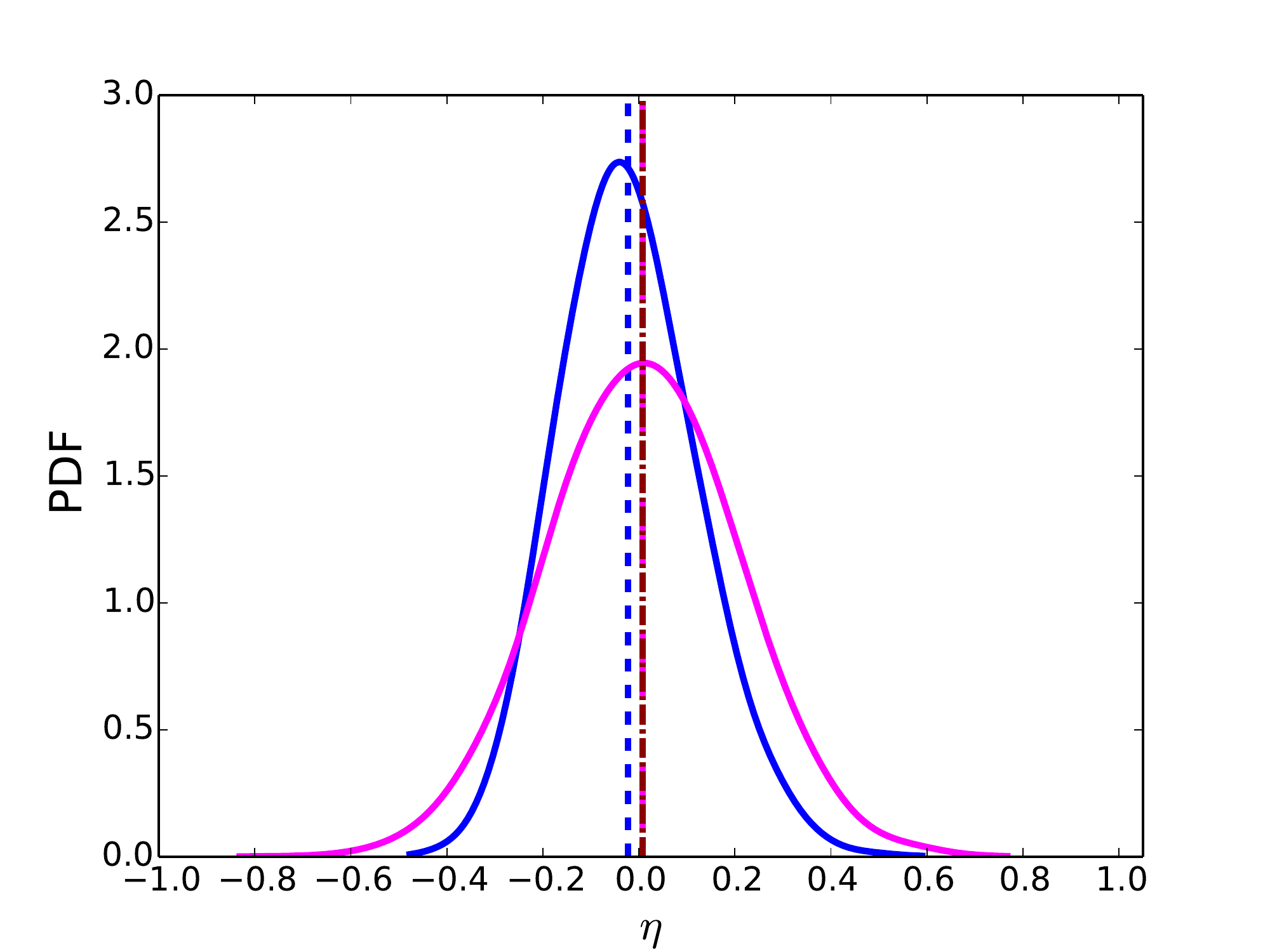}}
   \subfloat[PDF  of $\eta$, $\sigma = 0.6$]
   {\includegraphics[width=0.5\textwidth]{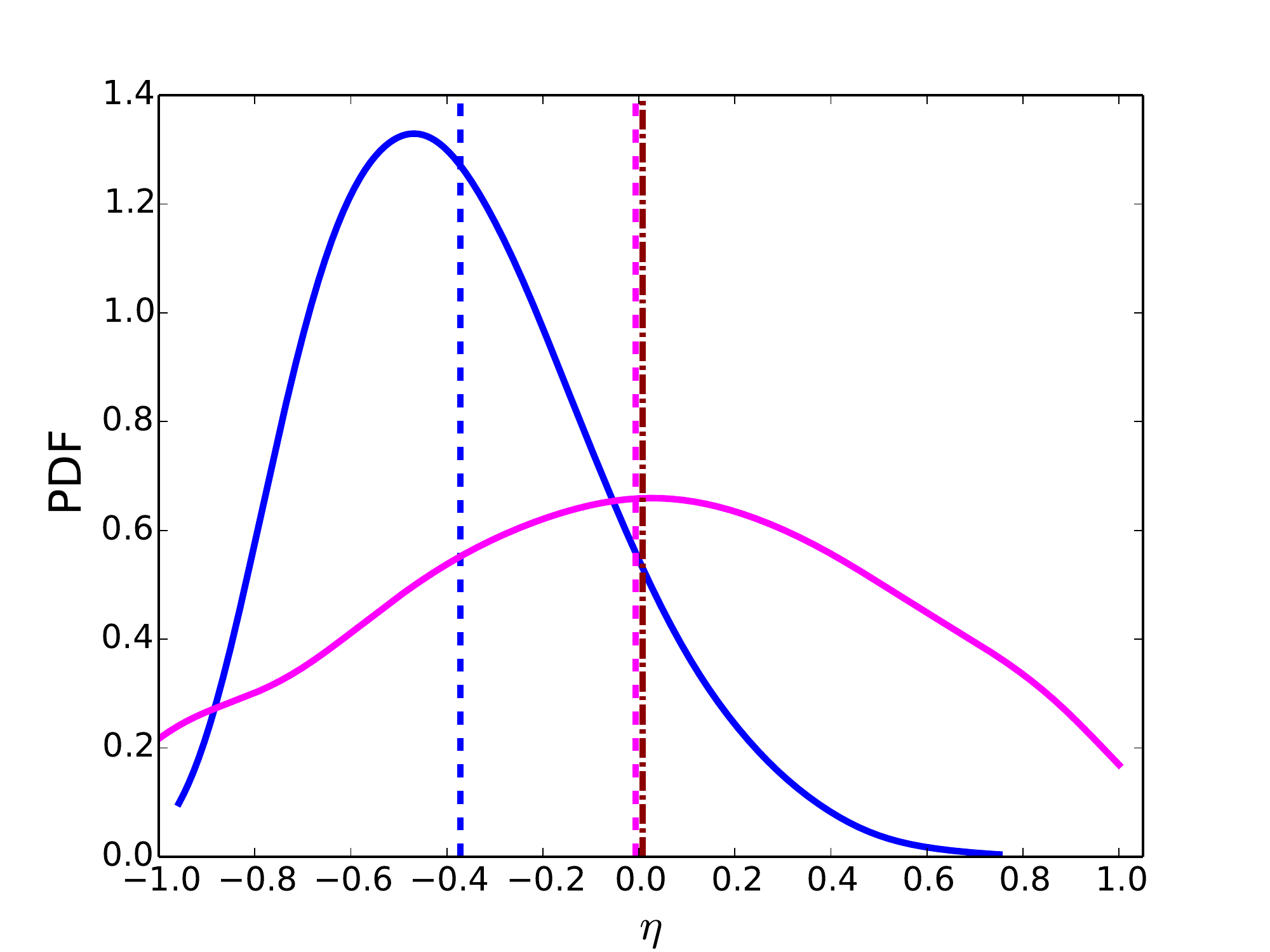}}\\   
   \caption{Distributions of the perturbed natural coordinates $\xi$ and $\eta$ for point A ($x/H = 2.0, y/H = 0.5$)
   located in recirculation region. The results from the physics-based approach and the RMT
   approach are compared in the same plot. Panels (a) and (b) show the comparison of $\xi$ 
   at perturbation variances of $\sigma = 0.2$ and $\sigma = 0.6$, respectively. 
   Panels (c) and (d) show the comparison of $\eta$ 
   at perturbation variances of $\sigma = 0.2$ and $\sigma = 0.6$, respectively.}
   \label{fig:comcon_XiEta_gen}
  \end{figure} 

The analysis above suggests that imposing Gaussian perturbation directly in Barycentric coordinates
(as oppose to the natural coordinates) leads to a distribution closer to maximum entropy. However, the perturbations
were imposed in natural coordinates by Xiao et al.~\cite{xiao-mfu} due to practical considerations.
Since the Barycentric coordinates $C_1$, $C_2$, and $C_3$ are correlated, and
the triangle boundary edges pose difficulties on the capping scheme, the natural coordinates $\xi$ and $\eta$ are
preferred for implementation purposes. In order to determine proper perturbations in $\xi$ and $\eta$ to obtain 
prior with maximum entropy in the physics-based approach, we also need to map the 
Barycentric coordinates to the natural coordinates in the RMT approach. 
With the samples of natural coordinates $\xi$ and $\eta$, their joint density can be estimated with 
Gaussian kernels. Figures~\ref{fig:contourNat_gen}a 
and~\ref{fig:contourNat_gen}b show the comparison of joint PDF contours obtained
by the two approaches with variances~$\sigma = 0.2$ and $\sigma = 0.6$, respectively.
Moreover, the comparisons of marginal distributions of $\xi$ and $\eta$
are also plotted in Fig.~\ref{fig:comcon_XiEta_gen}. When the perturbation
is small ($\sigma = 0.2$), the contours in both cases Phy1 and RM1 are elliptical, 
indicating the joint distributions are approximately Gaussian (Fig.~\ref{fig:comcon_XiEta_gen}a). 
However, a notable difference between the results of the two approaches lies on
the shapes of the contours. The elliptical contour in the RMT approach
is anisotropic with larger variance for $\xi$ but smaller for $\eta$.  In the physics-based
approach the contour is circular due to the artificial choice of the same perturbation variance
for both $\xi$ and $\eta$. Therefore, to achieve the prior with approximate maximum
entropy, the perturbation in $\xi$ should be larger than that in $\eta$ for this flow of concern. More details
can be found in their marginal distributions (Figs.~\ref{fig:comcon_XiEta_gen}a and~\ref{fig:comcon_XiEta_gen}c).
For both approaches with small perturbation ($\sigma = 0.2$) in Reynolds stress, 
the marginal distributions for $\xi$ and $\eta$ are Gaussian. 
With the RMT approach, the perturbation variance of $\xi$ is approximately twice as
large as that with the physics-based approach, while the
perturbation of $\eta$ is slightly smaller than that with the physics-based approach.
However, when the perturbation is large ($\sigma = 0.6$),
the joint distribution of $\xi$ and $\eta$ obtained by both approaches are
no longer Gaussian, and the density contours are influenced by the boundaries (Fig.~\ref{fig:contourNat_gen}b). 
Especially in case Phy2 the shape of the contour clearly follows the rectangular edges. 
Unlike the physics-based approach, the shape of the contour obtained with the RMT 
approach is less affected by the boundaries. Detailed comparisons are shown
by the corresponding marginal distributions. All the distributions are distorted compared to 
the Gaussian PDF (Gaussian density with the sample mean and variance). The distortions
in the physics-based approach are due to the fact that large number of unrealizable samples are 
capped onto the edges. However, in the RMT approach the sample
means of $\xi$ and $\eta$ significantly deviate from the baseline results. The mean
of $\xi$ moves towards the middle point ($\xi = 0$) of $\xi$ range, while the mean of $\eta$ moves
towards approximate one third ($\eta = -0.4$) of the $\eta$ range~$[-1, 1]$. This sample mean 
point ($\xi = 0.0$, $\eta = -0.4$) is close to the centroid of the triangle when 
mapped back to the Barycentric coordinates. Moreover, the distributions are approximately 
bounded Gaussian that satisfies the realizability constraint mathematically. 
The deviation of sample mean from the baseline when the perturbation is large can be interpreted intuitively.
Since the large perturbation of Reynolds stress implies less confidence on the baseline prediction,
it is reasonable to adjust the sample mean to the centroid of triangle to have a better
sample scattering over the entire triangle. The results shown in Figs.~\ref{fig:contourNat_gen}a,~\ref{fig:comcon_XiEta_gen}a,
and~\ref{fig:comcon_XiEta_gen}c suggest that, for a generic point away from the wall,
imposing small perturbations on $\xi$ and $\eta$ with Gaussian distributions 
lead to Reynolds stresses that are very close to the maximum entropy distribution. This
observation lends partial support to the choice of prior distributions made in~\cite{xiao-mfu}.
However, to achieve the maximum entropy prior, the perturbation variance of $\xi$ should be approximately 
twice as large as that of $\eta$. When the perturbation of Reynolds stress is large, to achieve maximum entropy
we should first shift the baseline $\xi$ and $\eta$ to the centroid of the Barycentric triangle,
and then perturb the shifted baseline results with bounded Gaussian distribution.  

  \begin{figure}[htbp]
  \centering
  \subfloat[$\sigma = 0.2$, location A]
   {\includegraphics[width=0.5\textwidth]{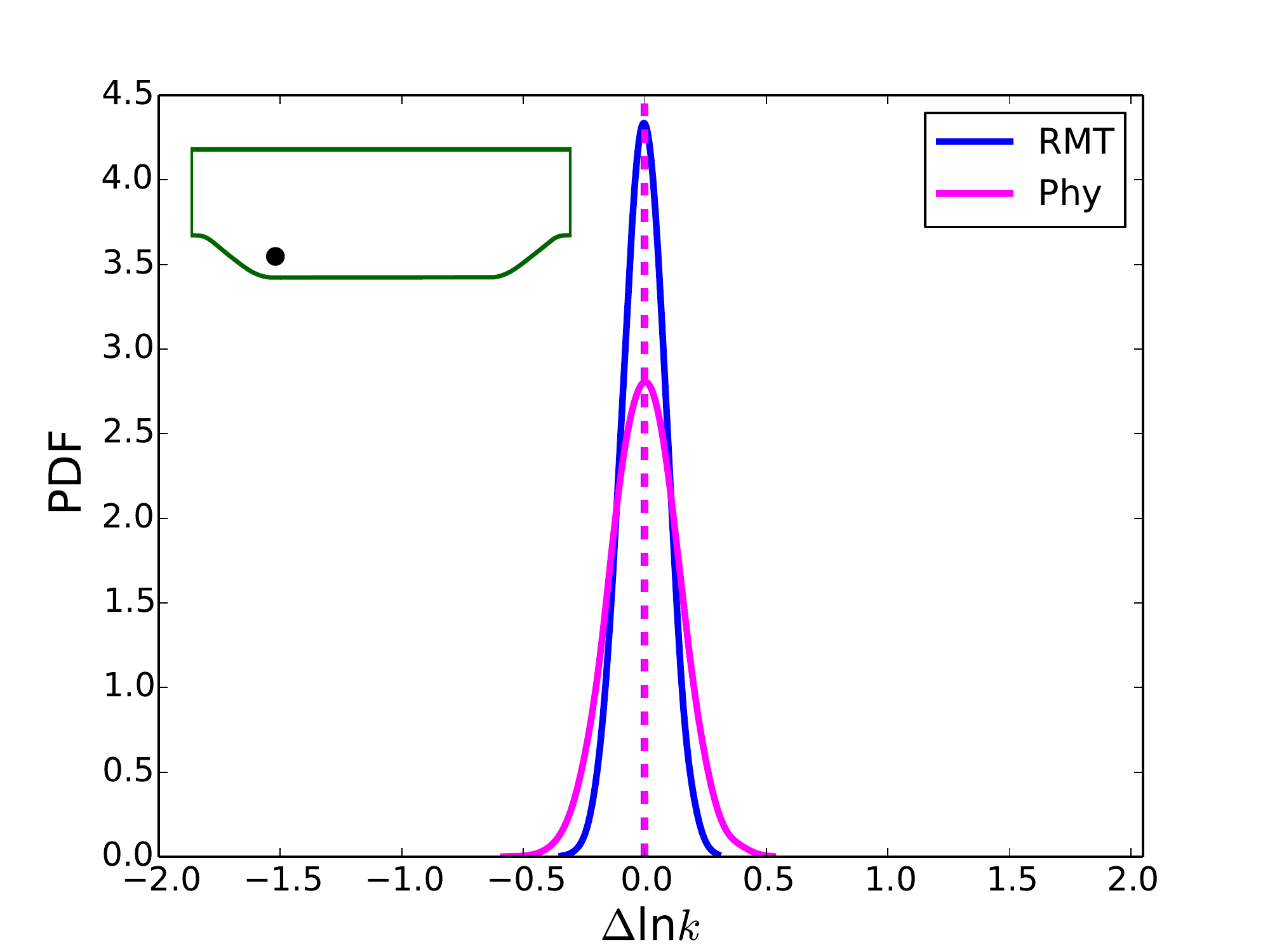}}
   \subfloat[$\sigma = 0.6$, location A]
   {\includegraphics[width=0.5\textwidth]{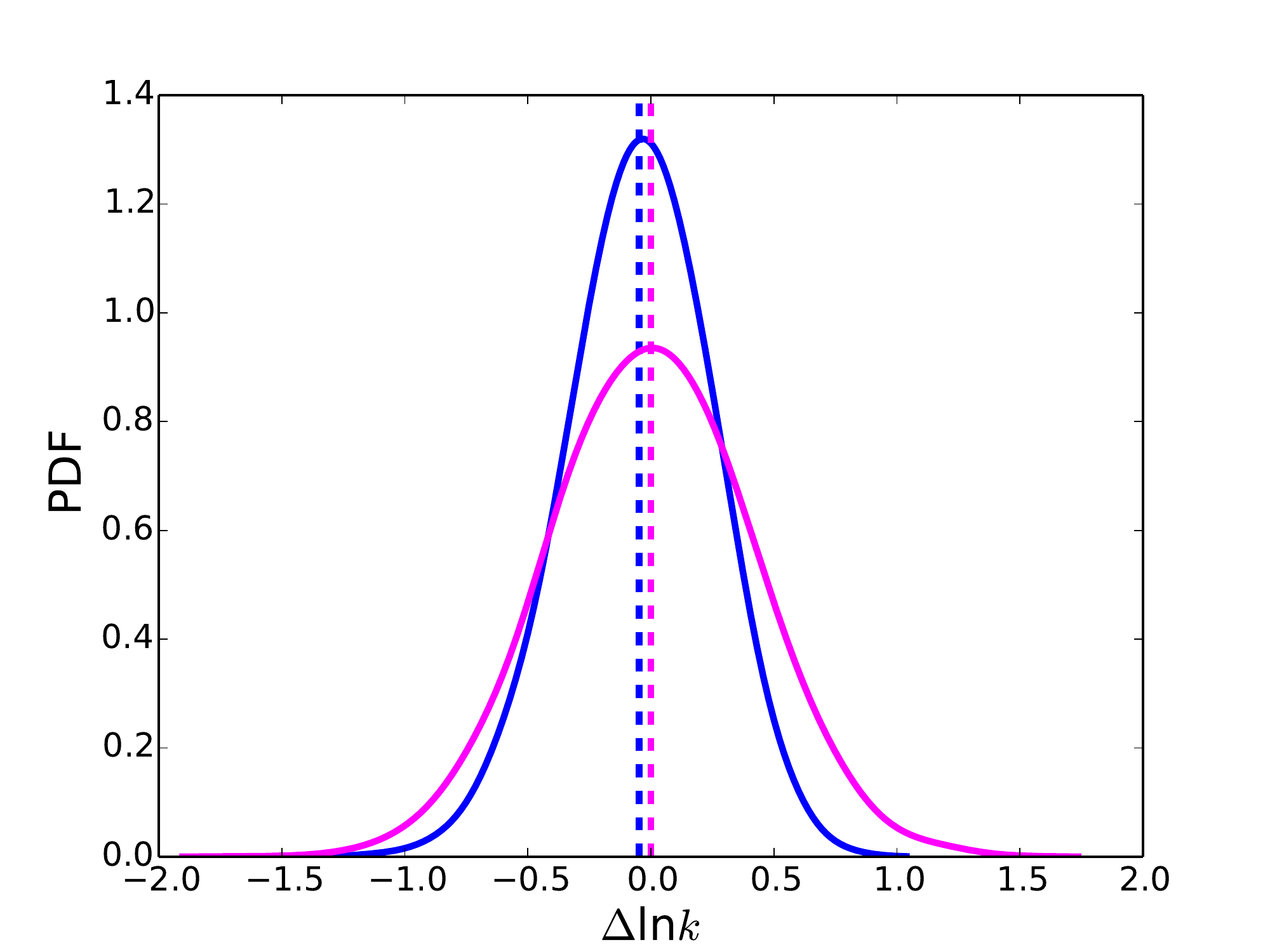}}\\
  \subfloat[$\sigma = 0.2$, location B]
   {\includegraphics[width=0.5\textwidth]{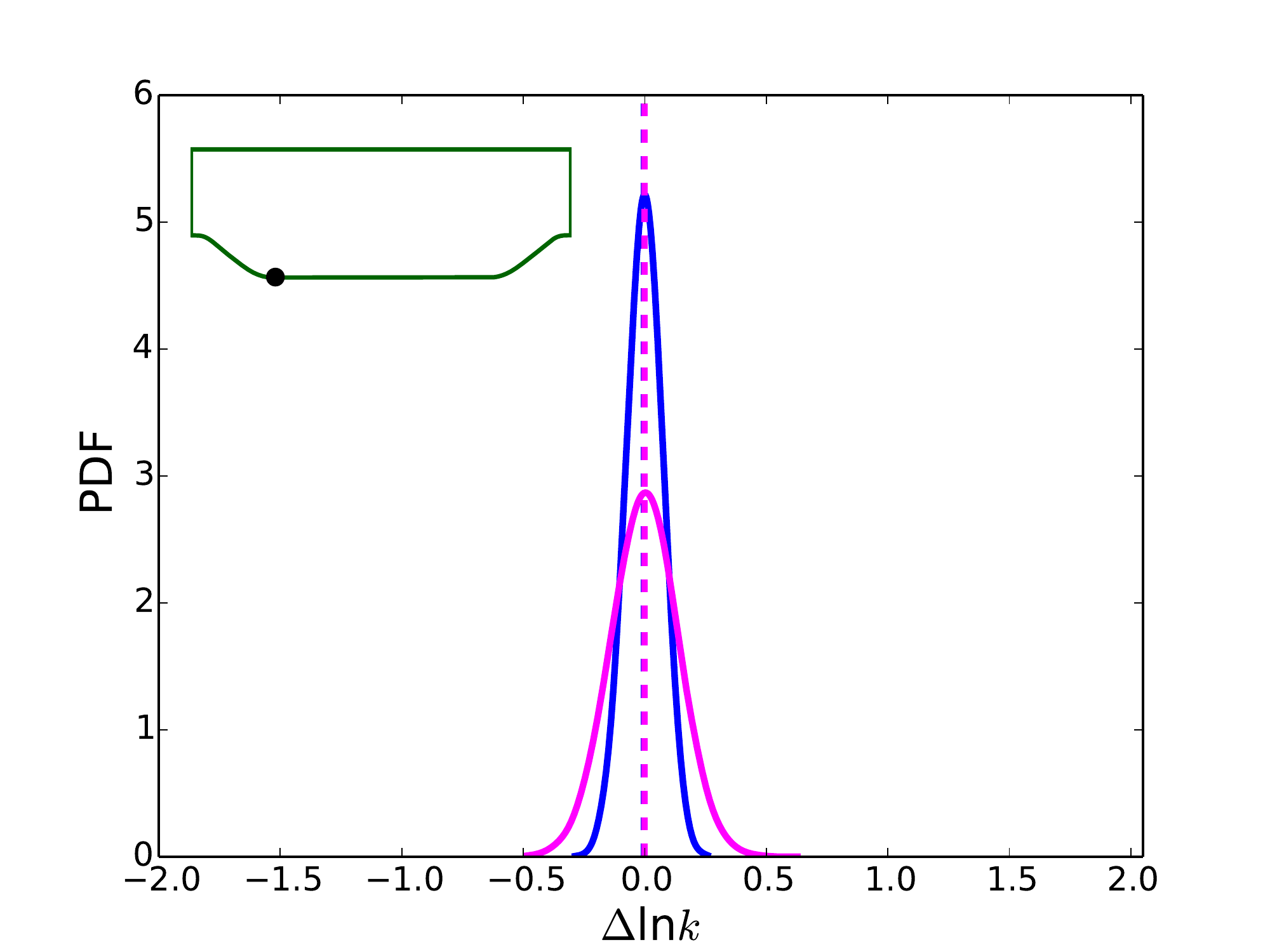}}
   \subfloat[$\sigma = 0.6$, location B]
   {\includegraphics[width=0.5\textwidth]{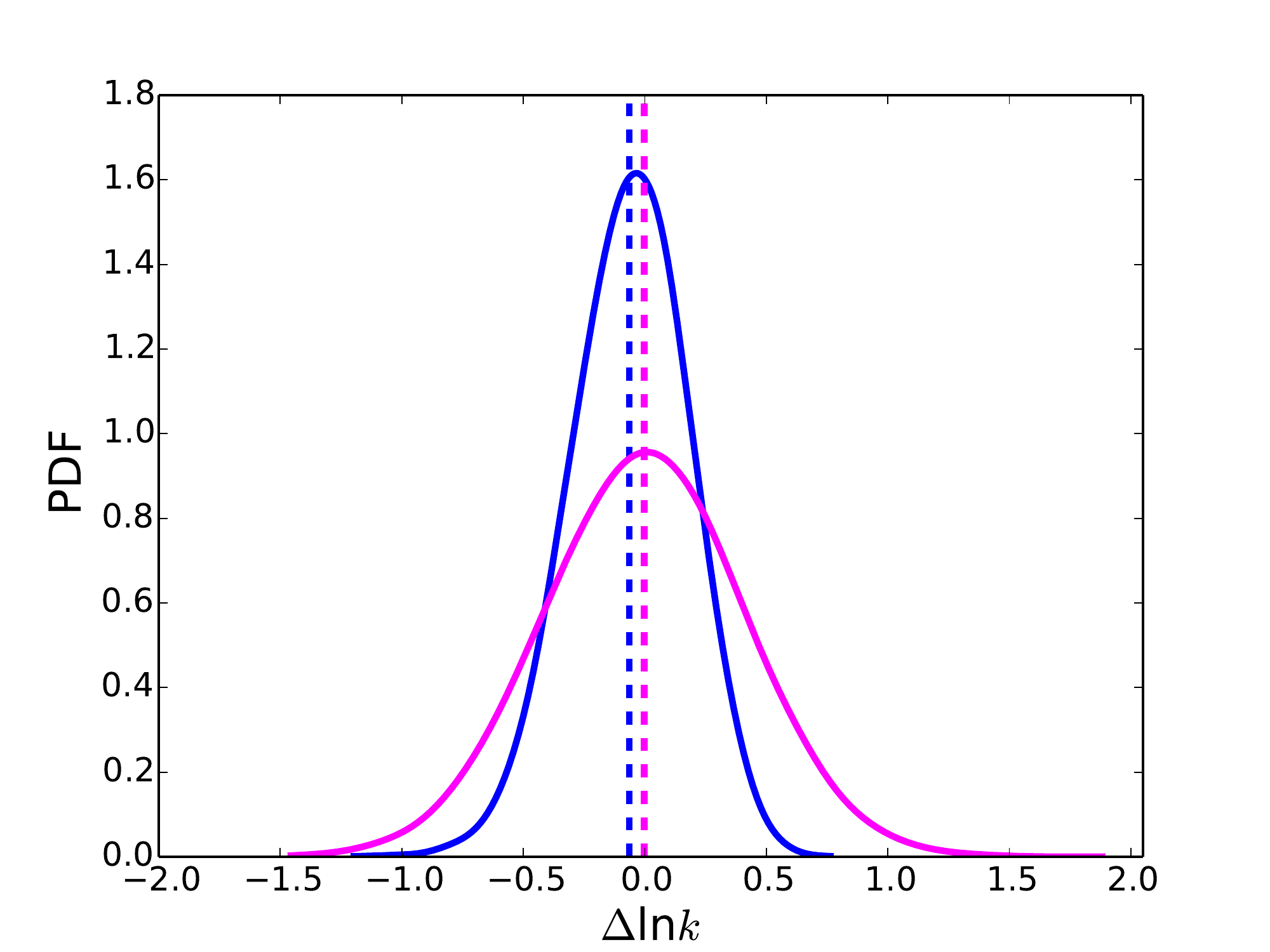}}      
   \caption{Distributions of the perturbations ($\Delta \log k$) in turbulence kinetic energy in 
   logarithmical scale. The upper two panels show the results at point A ($x/H = 2.0, y/H = 0.5$),
   while the lower two panels show the results at point B ($x/H = 2.0, y/H = 0.01$). 
   The results from physics-based approach and random matrix theoretic approach are compared
   in the same plot. The sample mean of the physics-based approach overlaps with that of the RMT approach 
   (denoted as blue (dark) dashed line and pink (grey) dashed line, respectively) }
   \label{fig:comcon_deltaK_gen}
  \end{figure}

In addition to the shape of Reynolds stress, its magnitude i.e., the turbulence kinetic energy (TKE), is also 
difficult to predict in RANS models. 
In the physics-based approach, the perturbations in turbulence kinetic energy are 
specified to be log-normally distributed. To evaluate if this specification is 
justified, we compared the TKE perturbations in logarithmic scale~$\Delta\log k$
obtained from the two approaches. The marginal distributions of~$\Delta \log k$ with
different perturbation levels ($\sigma = 0.2$ and $\sigma = 0.6$) at points A and B
are presented in Fig.~\ref{fig:comcon_deltaK_gen}. In both the physics-based and 
RMT approaches, the perturbations in TKE obey the log-normal
distribution, since all the PDFs of~$\Delta\log k$ shown in Fig.~\ref{fig:comcon_deltaK_gen}
are close to the Gaussian distributions. This conclusion is also true for the cases with
larger perturbations (cases Phy2 and RM2, $\sigma = 0.6$) as well. 
Therefore, introducing a log-normally distributed prior for TKE discrepancies
leads to a distribution of Reynolds stresses that is close to the one with maximum entropy. 
This lends support to the choice of prior in the physics-based approach~\cite{xiao-mfu}.
Another interesting observation in Fig.~\ref{fig:comcon_deltaK_gen} is that 
the spreading of~$\Delta\log k$ samples with the RMT approach is slightly smaller than
that with physics-based approach. As mentioned above, in the physics-based approach the same variance
field~$\sigma(x)$ is shared by the perturbations of six variables ($\xi$, $\eta$, $k$, $\varphi_1$, 
$\varphi_2$, $\varphi_3$)
due to the lack of prior knowledge. However, this assumption
is another constraint imposed in the physics-based approach. Based on the comparisons
in Fig.~\ref{fig:comcon_deltaK_gen}, we find that to achieve the maximum entropy,
the perturbation variance for each parameter should be different. 
It suggests that a relatively smaller variance (approximate 50\% of that for shape parameter) is
proper for the perturbation of TKE in logarithmic scale for this flow of concern. 

\begin{figure}[htbp]
  \centering
  \subfloat[$\sigma = 0.2$, at location A]
   {\includegraphics[width=0.5\textwidth]{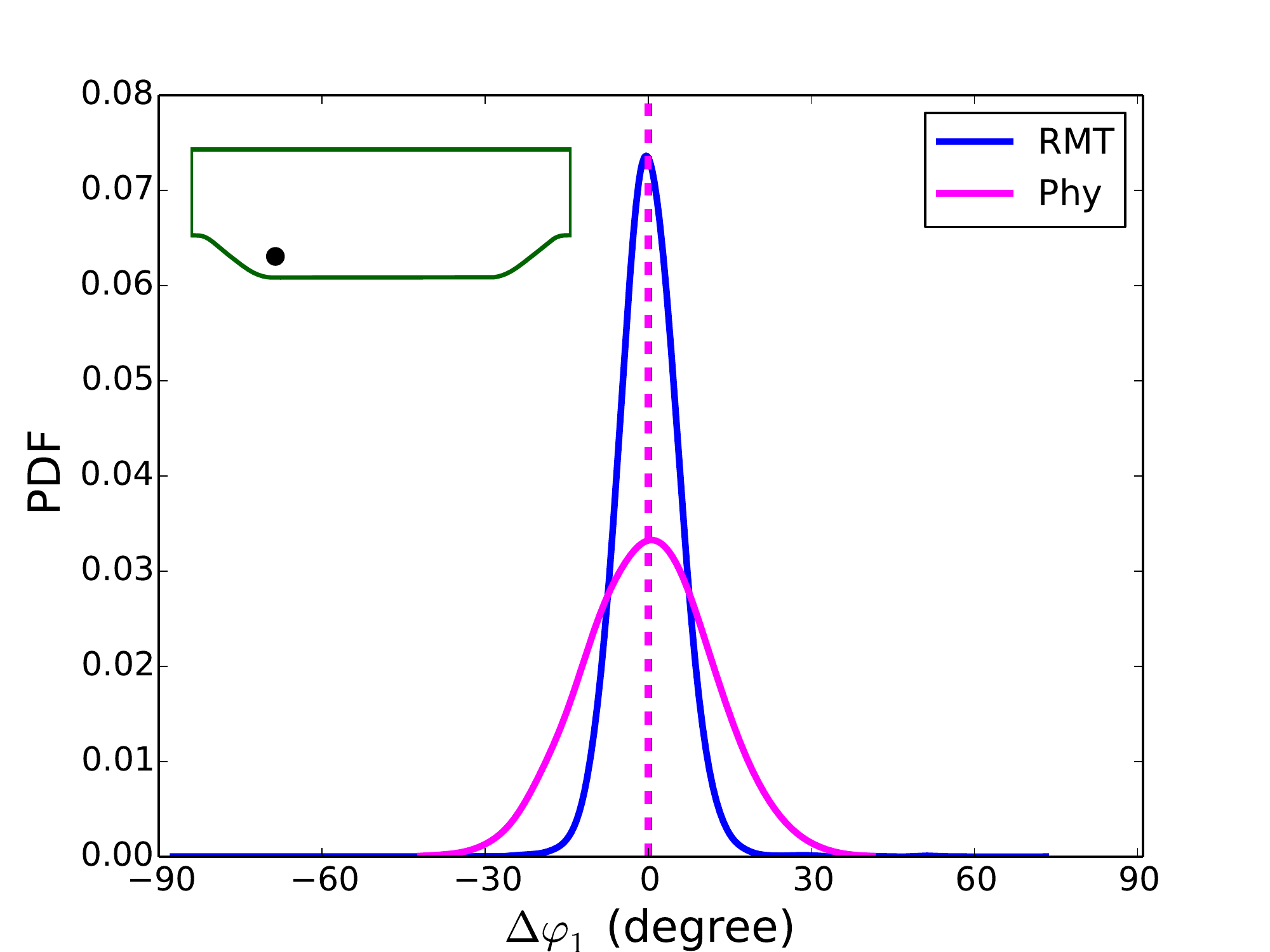}}
   \subfloat[$\sigma = 0.6$, at location A]
   {\includegraphics[width=0.5\textwidth]{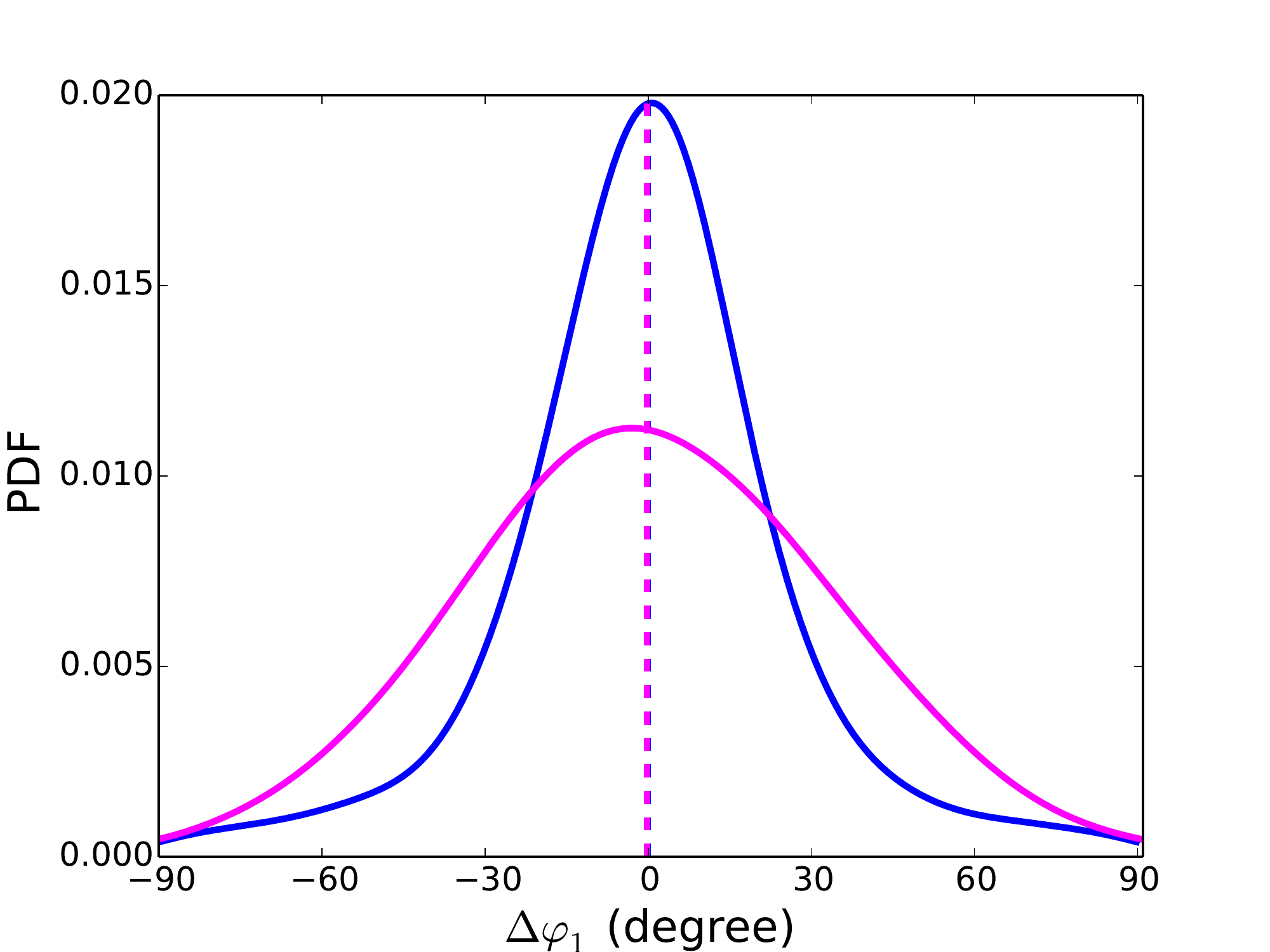}}\\
  \subfloat[$\sigma = 0.2$, at location B]
   {\includegraphics[width=0.5\textwidth]{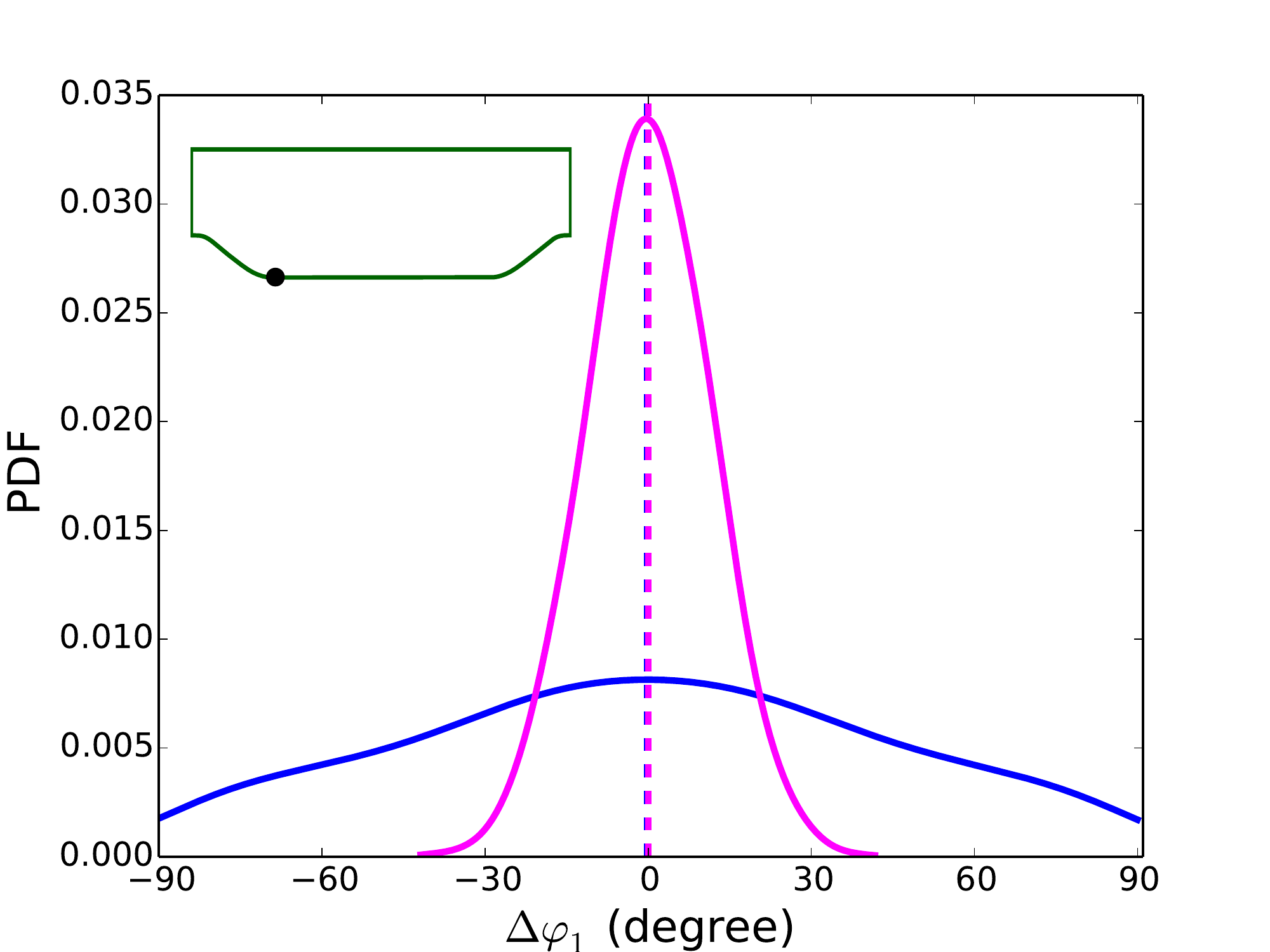}}
   \subfloat[$\sigma = 0.6$, at location B]
   {\includegraphics[width=0.5\textwidth]{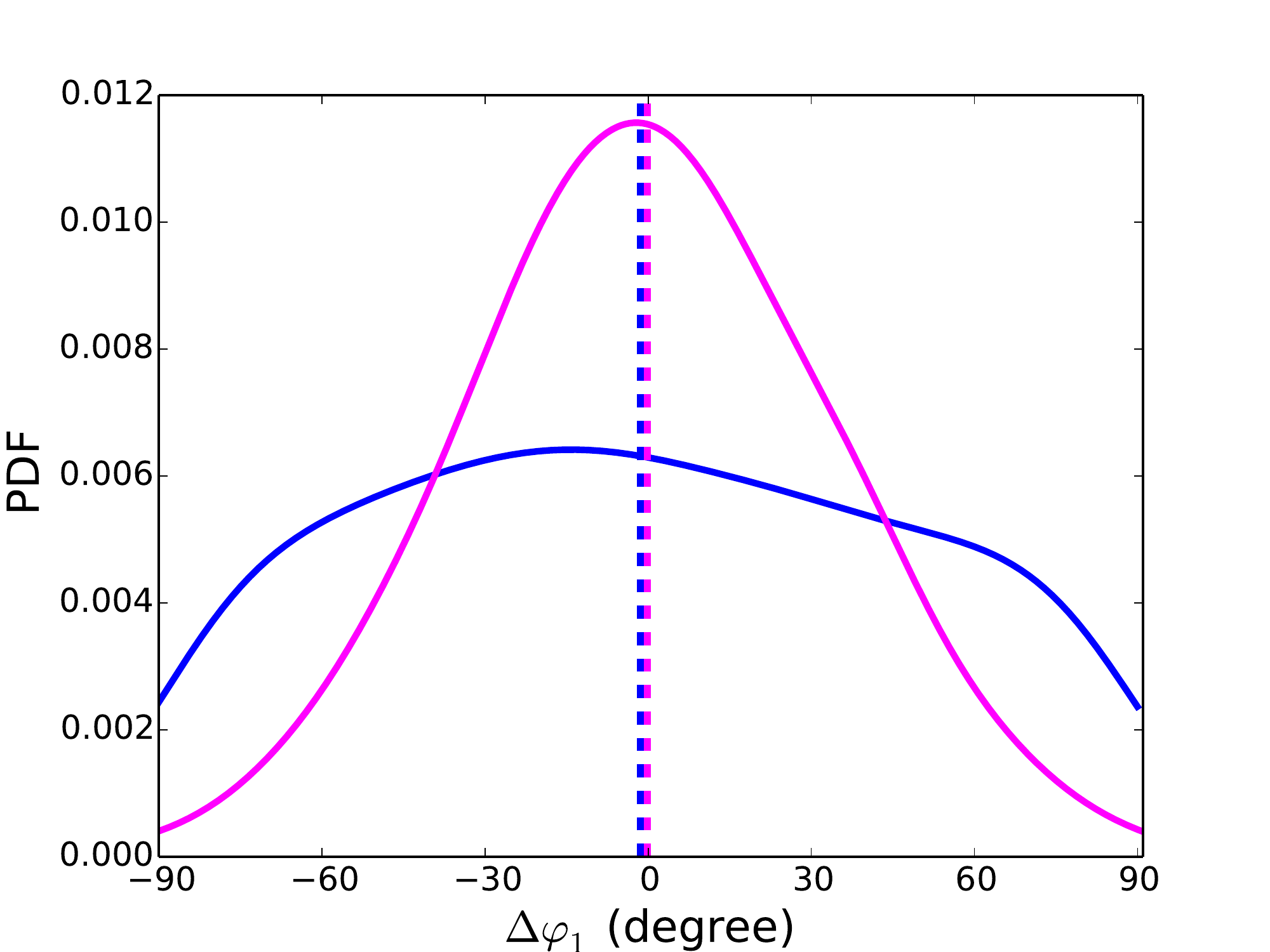}}   
   \caption{ 
   Distributions of the perturbations ($\Delta \varphi_1$) in orientation. 
   The upper two panels show the results at point A ($x/H = 2.0, y/H = 0.5$),
   while the lower two panels show the results at point B ($x/H = 2.0, y/H = 0.01$). 
   The results from Physics-based approach and random matrix theoretic approach are compared
   in the same plot. The sample mean of the physics-based approach overlaps with that of the RMT approach 
   (denoted as blue (dark) dashed line and pink (grey) dashed line, respectively) }
   \label{fig:comcon_deltaVA}
  \end{figure} 

While the orientations of the Reynolds stresses have not been perturbed in~\cite{xiao-mfu}, 
in this study they are perturbed with the same variance~$\sigma$ as that of other variables. 
However, if or how large the orientation should be perturbed are model choices in the physics-based approach. 
Without further physical information, it is difficult to determine the variance of perturbations in angles. 
To examine this issue, the marginal distributions of angle discrepancies obtained from both approaches are compared. 
Fig.~\ref{fig:comcon_deltaVA} shows the comparison of PDF of~$\Delta\varphi_1$ with 
different perturbation magnitudes at the points A and B.
The perturbations in $\varphi_2$ and $\varphi_3$ have similar characteristics as that of $\varphi_1$, and
thus they are omitted for brevity. We can see that to achieve the maximum entropy the orientation of the tensor
should also be perturbed. Similarly to the perturbations in TKE, 
the sample mean of~$\Delta\varphi_1$ is zero for all cases. 
This indicates that the sample means in tensor magnitude and orientation
are the same as those of baseline RANS prediction, since there are no physical constraints on TKE and Euler
angles (except for the range $[-\pi, \pi]$ specified in the definition).
In the physics-based approach, the PDFs are Gaussian for all cases, since
Gaussian random fields are employed to model the discrepancies in angles.  
For the generic point A, the distribution of~$\Delta\varphi_1$ obtained
from the RMT approach is also close to Gaussian when the perturbation is small 
(Fig.~\ref{fig:comcon_deltaVA}a). However, the scattering of 
samples is smaller than what we specified in the physics-based approach, 
and most of the perturbations are within $\pm20$ degrees.  As the perturbation magnitude is enlarged, 
the PDF obtained in the RMT approach is peaked compared
to the Gaussian distribution (Fig.~\ref{fig:comcon_deltaVA}b).
Consequently, most samples are with small perturbations (smaller than $30$ degrees). 
However, for the point B located close to the wall, the variances of~$\Delta\varphi_1$ in 
cases RM1 and RM2 are much larger, even though the perturbation of tensor is small
(Fig.~\ref{fig:comcon_deltaVA}c). In the RMT approach 
the distribution of~$\Delta\varphi_1$ is flatter than the corresponding Gaussian distribution, 
and most of the perturbations~$\Delta\varphi_1$ are larger than $60$ degrees. 
Especially when the dispersion parameter is large (i.e., $\delta(x)$ corresponding to $\sigma = 0.6$), the
PDF of~$\Delta\varphi_1$ significantly deviates from the Gaussian distribution (Fig.~\ref{fig:comcon_deltaVA}d). 
The observations above imply that, to obtain a maximum entropy prior, the perturbations variance
in orientation should be spatially non-stationary. For this specific flow, 
smaller perturbations (with 30 degrees) should be used for
a generic point away from the way, while larger perturbations are appropriate for the near wall locations.

The results shown above suggest that, when the RANS predictions are relatively reliable with small
perturbations needed, using normally distributed perturbations for each of the six physical
variables is a good choice to obtain the Reynolds stress prior that is close to the one with maximum
entropy. This observation can be related to the case of a scalar random variable.  With the
constraints of a specified mean and variance, the maximum entropy distribution of a scalar random
variable is the Gaussian distribution~\cite{park2009maximum}.  That is, by choosing Gaussian
distributions for the discrepancies of Reynolds stresses in the physics-based approach, the maximum
entropy distribution is achieved for each individual physical variables (magnitude, shape,
orientation). The variance of each variable is, however, chosen by the user and is thus a modeling
choice.  In contrast, in the RMT approach the maximum entropy is achieved for the distribution of
the Reynolds tensor.  Consequently, the relative magnitude of the variance for each variable is
implied based on the maximum entropy principle and is not a modeling choice of the user.  However,
it is worth pointing out that, although we use the results of the RMT approach as the golden
standard for gauge the physics-based approach, the correlation structure (see Eq.~\ref{eq:Lx}) and
corresponding Karhunen-Loeve modes used to represent the random field are still modeling choices,
which also may introduce artificial constraints. In summary, one can consider that in the RMT
approach, maximum entropy is achieved for the pointwise distribution of the Reynolds stresses 
at each location $x$, but not necessarily for the random field
$[\mathbf{R}(x)]$. In the physics based approach, maximum entropy is achieved for the pointwise
distribution for individual physical variables ($k$, $\xi$, $\eta$ etc.) but not necessarily for the
Reynolds stress tensor $[\mathbf{R}]$ or the field $[\mathbf{R}(x)]$.

\section{Conclusion}
\label{sec:con}
Quantification of the uncertainties originating from the modeled Reynolds stresses is crucial when
the RANS simulations are applied in the decision-making process.  One of the challenges in current
form of the physics-based model-form uncertainty quantification framework is to specify proper
priors for the physical variables (shape, magnitude, and orientation). It is difficult to determine
if or how much additional information is introduced into the priors specified in the physics-based
approach. To evaluate the priors and gain insights on proper specification of the priors, the random
matrix theoretic approach with the maximum entropy principle is used in this work. By comparing the
distributions of shape, magnitude, and orientation variables obtained from the two approaches, we
find some useful guidelines of prior specification for these wall-bounded flows with separations.
For the shape parameters, the Gaussian distributed perturbations in Barycentric coordinate is better
to achieve maximum entropy. The mapping between the Barycentric coordinate to the natural coordinate
may introduces some artificial constraints, especially when the RANS prediction is less reliable
(large perturbation needed).  For the turbulence kinetic energy, introducing log-normally
distributed perturbations leads to a distribution of Reynolds stresses that is close to the one with
the maximum entropy.  This observation lends support to the choice of prior in~\cite{xiao-mfu}.
Moreover, different variance fields for the perturbations of the Reynolds stress shape parameters
$\xi$, $\eta$, and the magnitude (turbulence kinetic energy) $k$ are required to achieve maximum
entropy. Specifically, for the flow over periodic hills examined in this study the variance of $\xi$
should be larger than that of $\eta$, and the perturbation of~$\log k$ should be relatively
small. Finally, it suggests that the uncertainties should be also injected in the orientation of
Reynolds stresses represented by Euler angles.  For a generic location away from the wall, the
perturbations of Euler angles should be small, while larger perturbations should be added for the
near wall locations.  The conclusion can be used as a guidance for the objective prior specification
in the physics-based, Bayesian uncertainty quantification framework.




\appendix

\section{Summary of Algorithms of The Physics-Based Approach}
\label{apped:phy}
\begin{enumerate}

\item Decomposition to physically meaningful dimension and expansion of given marginal distributions
\begin{enumerate}[label*=\arabic*.]
\item Perform the baseline RANS simulation to obtain the baseline (mean)
  Reynolds stress $[\underline{R}]$.
\item Perform the transformation $[\underline{R}] \mapsto ( \tilde{\xi}^{rans}, \,
  \tilde{\eta}^{rans}, \,  \tilde{k}^{rans}, \, \tilde{\varphi}_1^{rans}, \, \tilde{\varphi}_2^{rans}, \, \tilde{\varphi}_3^{rans})$.  
\item Compute Karhunen--Loeve expansion to obtain basis set $\{\phi_{\alpha}(x)\}_{\alpha=1}^{N_{kl}}$, where $N_{kl}$ is the number of
  modes retained.  
\end{enumerate}

\item Sampling and reconstruction of physical variable fields for Reynolds stresses:
\begin{enumerate}[label*=\arabic*.]
\item Sample six independent coefficient vectors $\{ \bs{\omega}_{\beta} \}_{\beta=1}^N$ for the six discrepancy fields 
(i.e., $\Delta\xi, \Delta\eta, \Delta\log{k}, \Delta\varphi_1, \Delta\varphi_2$, and $\Delta\varphi_3$  ), where $N$ is the sample size.
\item Reconstruct the six discrepancy fields with the six independent coefficient samples~$\{ \bs{\omega}_{\beta} \}_{\beta=1}^N$
	and Karhunen--Loeve modes. Note that the variance field $\sigma(x)$ is the same for the six random fields.
\item Obtained samples of Reynolds stress field $[\mathbf{R}]$ via mapping 
$( \xi, \, \eta, \, k, \, \varphi_1, \, \varphi_2, \, \varphi_3) \mapsto  [R]$
\end{enumerate}

\item Propagation the Reynolds stress to QoIs via RANS equations
\begin{enumerate}[label*=\arabic*]
\item Use the obtained sampled Reynolds stress to velocity and other QoIs by solving the RANS
  equations.
\item Post-process the obtained velocity and QoI samples to obtain statistical moments.
\end{enumerate}

\end{enumerate}

\section{Summary of Algorithms of The RMT Approach}
\label{apped:rmt}
Given the mean Reynolds stress field $[\underline{R}(x)]$ (e.g., from RANS-predicted results)
along with the correlation function structure of the random upper triangle matrix field 
$[\mathbf{L}](x)$, the following procedure is performed:

\begin{enumerate}
\item  Expansion of given marginal distributions and covariances kernels:
  \begin{enumerate}[label*=\arabic*.]
  \item Perform the Cholesky factorization of the mean Reynolds stresses $[\underline{R}]$ at each
    cell as $[\underline{R}] = [\underline{L}_R]^T [\underline{L}_R]$, 
    which yields field $\underline{L}_R(x)$ of upper triangular matrices.
  \item Perform Karhunen--Loeve expansion for the kernel function by solving the Fredholm equation to
    obtain eigenmodes.
  \item For off-diagonal terms of matrix $[\mathbf{L}]$, perform polynomial expansion (PCE) of the 
  Gamma marginal PDF at each cell. PCE Coefficients $U_\beta$ are obtained from
 \begin{equation}
  \label{eq:U_i}
  U_\beta  = \frac{\langle \mathbf{u} {\Psi}_\beta \rangle}{\langle {\Psi}_\beta^2 \rangle} \\
   =  \frac{1}{\langle {\Psi}_\beta^2 \rangle} \int_{\Omega}
   F_{\mathbf{u}}^{-1}[F_{\mathbf{w}}(w)] \; {\Psi}_\beta({w})  \; p_{\text{\tiny \textbf{w}}}(w) dw, 
\end{equation}
where $\langle \boldsymbol{\Psi}_\beta^2 \rangle$ is
the variance of $i$\textsuperscript{th} order polynomial of standard Gaussian random variable
$\mathbf{w}$; $F_w(\mathbf{w})$ and $p_{\text{\tiny \textbf{w}}}$ are the cumulative distribution
function (CFD) and PDF, respectively, of $\mathbf{w}$; $\Omega$ is the sample space of $\mathbf{w}$;
$F_{\mathbf{u}}$ and $F_{\mathbf{u}}^{-1}$ are the CDF and its inverse, respectively, of random
variable $\mathbf{u}$. The index $\beta$ is from $1$ to $N_p$, and $N_p$ is the number of polynomials
retained in the expansion. 
 \end{enumerate}

\item Sampling and reconstruction of random matrix fields for Reynolds stresses:

\begin{enumerate}[label*=\arabic*.]
\item For each element $\mathbf{L}_{i j}$ of the random matrix field $[\mathbf{L}]$, independently
  draw $N_{kl}$ sample from the standard Gaussian distribution $\omega_{i j, \alpha}$ where
  $\alpha = 1, \cdots, N_{kl}$, e.g., with random sampling or Latin hypercube sampling
  method.

\item Synthesize realizations of the off-diagonal terms based on Karhunen--Loeve expansion:
  \begin{align}
    \mathbf{w}_{ij}(x) & = \sum_{\alpha=1}^{N_{\textrm{KL}}} \,
    \phi_\alpha(x) \;    \boldsymbol{\omega}_\alpha   \quad \textrm{with} \quad  i < j \notag \\ 
    \mathbf{L}_{i j}(x) & = \sigma_d \mathbf{w}_{ij}(x) \notag
  \end{align}

\item Synthesize the realizations of the diagonal terms based on Karhunen--Loeve and PCE expansions:
  \begin{equation}
    \mathbf{u}_{i}(x) = \sum_{\beta = 0}^{N_p} U_\beta(x) \Psi_\beta(\mathbf{w}_{ii}(x))   \notag
  \end{equation}
  where the Gaussian random field sample $\mathbf{w}_{ii}(x)$ obtained in the previous step is used.
\item Synthesize the diagonal terms of matrix $[\mathbf{L}]$ from $\mathbf{L}_{i i}(x) = \sigma_d
  \sqrt{2\mathbf{u}_{i}}$, where $i = 1, 2, 3$.
\item Reconstruct random normalized matrix $[\mathbf{G}]$ from $[\mathbf{G}] = [\mathbf{L}]^T [\mathbf{L}]$ and then
  reconstruct the Reynolds stress tensor $[\mathbf{R}]$ from $[\mathbf{R}] = [\underline{L}_R]^T [\mathbf{G}]
  [\underline{L}_R]$.
\end{enumerate}

\item Propagation the Reynolds stress field through the RANS solver to obtain velocities and
    other QoIs:

\begin{enumerate}[label*=\arabic*]
\item Use the obtained sampled Reynolds stress to velocity and other QoIs by solving the RANS
  equations.
\item Post-process the obtained velocity and QoI samples to obtain statistical moments.
\end{enumerate}

\end{enumerate}

\section{Nomenclature}
\label{app:notation}
\begin{tabbing}
  000000000\= this is definition\kill 
   {\emph{Subscripts/Superscripts}}{}\\
  $i$, $j$ \> tensor indices ($i, j = 1, 2, 3$); repeated indices does not imply summation \\
  $\alpha$, $\beta$ \> indices for terms in Karhunen--Loeve  and  polynomial chaos expansions \\ 
  $l$ \> general index (modes etc.) \\
  \\
 \emph{Sets, operators, and decorative symbols} \>  {} \\
  $\mathbb{R}^{+}$ \> the set of all real positive number \\
  $\mathbb{E}\{ \cdot \}$ \> expectation of a random variable \\
  $\mathbb{M}_d^{s}$ \> the set of all $d \times d$ symmetric matrices \\
  $\mathbb{M}_d^{+}$ \> the set of all $d \times d$ symmetric, positive definite matrices \\
  $\mathbb{M}_d^{+0}$ \> the set of all $d\times d$ symmetric, positive semi-definite matrices \\
  $\operatorname{tr}$ \> trace of a matrix \\
  $\operatorname{det}$ \> determinant of a matrix \\
  $[\cdot]$ \>  matrix \\
  $\underline{\Box}$ \> mean value of variable $\Box$ \\
  $\bullet$ \> inner product of two (determistic) vectors \\
  $\langle \cdot \rangle$ \> ensemble average/expectation of a random variable \\
  $\| \cdot \|_F$  \> Frobenius norm \\
  $\sum$ \> summation \\
  $\prod$ \> product \\
  \\
  \emph{Roman letters} \>  {} \\
  $[A]$  \> anisotropy tensor of the Reynolds stress \\
  $C$ \> Barycentric coordinates\\
  $d$ \> dimension of matrices ($d=3$ implied unless noted otherwise) \\
  $D_{KL}$ \> Kullbck-Leibler divergence \\  
  $\vec{e}_1$, $\vec{e}_2$, $\vec{e}_3$  \> orthonormal eigenvectors of Reynolds stress anisotropy $[A]$ \\ 
  $[E]$  \> orthonormal eigenvectors of Reynolds stress anisotropy $[A]$ \\
  $F$  \> cumulative distribution function \\
  $[\mathbf{G}], [G]$ \> a positive definite matrix with identity matrix $[I]$ as its mean \\ 
  $[I]$ \>  identity matrix \\
  $k$ \>  turbulence kinetic energy \\
  $K$ \>  covariance kernel \\
  $[\mathbf{L}], [L]$ \> an upper triangle matrix (e.g., obtained from Cholesky factorization) \\
  $l$ \> length scale of Gaussian process \\
  $\mathbf{N}$  \> Gaussian normal distribution \\
  $N_p$ \> order of polynomial in the PCE expansion \\
  $N_{kl}$ \> number of modes in KL expansion \\
  $p, q$ \> probability density function \\
  $[\mathbf{R}]$, $[R]$ \> (negative of) Reynolds stress tensor with $R_{ij} = \langle \mathbf{v}_i' \mathbf{v}_j' \rangle$  \\
  $\mathbf{v}_i'$ \>  the $i$\textsuperscript{th} component of the fluctuation velocity (random
  variable) \\
  $\mathbf{u}$ \> Gamma random variable\ used to define the diagonal terms of the Reynolds stress $\mathbf{L}_{\alpha \alpha}$  \\
  $U$ \>  coefficients for the polynomial \\
  $\mathbf{w}$ \> standard Gaussian random variable \\
  $x$ \> spatial coordinate index\\
  \\
  
  {\emph{Greek letters}}{}\\
  $\delta$ \> dispersion parameter (uncertainty of the random matrix) \\
  $\sigma$ \> variance of Gaussian process\\
 $\vartheta$ \> parameter \\
  $\rho$ \> correlation\\  
  $\xi$, $\eta$ \> natural coordinates\\
  $\Lambda$, $\lambda$ \> eigenvalues of anisotropy $[A]$ \\ 
  $\varphi$  \> Euler angle of Reynolds stress tensor\\
  $\tilde{\omega}, \phi$ \> eigenvalues and basis functions obtained from Karhunen-Loeve expansion \\
  $\Psi_j$ \> $j$\textsuperscript{th} order Hermite polynomial of standard Gaussian random variable $\mathbf{w}$
 \end{tabbing}


\end{document}